\documentclass[aps,prd,preprint,nofootinbib,superscriptaddress,11pt]{revtex4-2}

\usepackage[T1]{fontenc}
\usepackage[utf8]{inputenc}
\usepackage{lmodern}

\usepackage{amsmath}
\usepackage{amssymb}
\usepackage{amsfonts}
\usepackage{amsthm}

\usepackage{mathtools}
\usepackage{mathrsfs}
\usepackage{bm}
\usepackage{tikz}
\usetikzlibrary{decorations.pathmorphing}

\usepackage{graphicx}
\usepackage{xcolor}

\usepackage{booktabs}
\usepackage{multirow}

\usepackage[
colorlinks=true,
linkcolor=blue,
citecolor=blue,
urlcolor=blue
]{hyperref}

\newcommand{\ii}{\mathrm i}

\newcommand{\ad}{\operatorname{ad}}
\providecommand{\eth}{\text{\dh}}
\renewcommand{\eth}{\text{\dh}}

\begin{document}

\title{Perturbative Reconstruction of Self-Adjoint Generators from
Bosonic Canonical Commutation Relations:
Application to the Null-Surface Formulation}
\author{C.~N.~Kozameh}
	\email{carlos.kozameh@unc.edu.ar}
	\affiliation{FaMAF, Universidad Nacional de C\'ordoba,
		5000 C\'ordoba, Argentina}
	
	\date{\today}

\begin{abstract}
We study the inverse adjoint problem for a perturbative bosonic outgoing
map.  Our main result is a constructive integrability theorem: within the
polynomial bosonic CCR algebra, preservation of the canonical commutation
relations order by order is sufficient for the explicit recursive
reconstruction, to all perturbative orders, of a formal self-adjoint series
\(\delta T(\varepsilon)=\sum_{m\geq1}\varepsilon^m\delta T_m\) whose
exponential implements the outgoing map by adjoint conjugation.  No
independent ansatz for the scattering generator is required.  At each
order, the BCH contribution fixed by previously reconstructed generators
is subtracted from \(\delta a_n^{\rm out}\); the outgoing CCRs then force
the residual to satisfy the homogeneous inverse-commutator conditions,
from which the new coefficient \(\delta T_{n-1}\) follows in closed form.
The construction also provides an independent order-by-order consistency
test of perturbatively computed outgoing operators.
We apply this result to graviton scattering in the null-surface formulation
(NSF).  In NSF the variables \(x^a\) label the four-dimensional solution
space of the cut equation; spacetime geometry is reconstructed on this
solution space rather than introduced as an external background.  In the
sector studied here, the reconstructed metric on that solution space is
specialized to the flat metric \(\eta_{ab}\), while the associated null
generators, affine parameter, and cone measure are kept at their
corresponding flat-space values.  The NSF outgoing map preserves
the Ashtekar radiative canonical commutation relations through
\(\delta a_3\), namely through order \(\varepsilon^2\).  At the
\(\delta a_3\) level,
the CCRs select the symmetric ordering \(c=1\), whose induced linear
contraction cancels the double-contraction contribution from
\([\delta a_2,\delta a_2^\dagger]\).  The reconstruction theorem then
yields a formal perturbative unitary implementation through this order.
The corresponding self-adjoint generators \(\delta T_1\) and \(\delta T_2\)
are reconstructed explicitly.  For the direct \(2\to2\) matrix element,
the same canonical ordering also fixes the complete
\(\delta a_3\delta a_3\) decomposition into cubic--cubic,
linear--cubic, cubic--linear, and linear--linear sectors, thereby separating
the irreducible loop from the one-particle-reducible and factorized
contributions.  Exponentiating the reconstructed generators organizes the
same scattering map into point vertices, canonical
contractions through on-shell intermediate states, and the associated loop
topologies.
\end{abstract}

\maketitle

\section{Introduction}
\label{sec:introduction}

The quantum NSF program was initiated in
Ref.~\cite{KozamehZapataAltuna2025} by quantizing the classical nonlinear
scattering construction developed in
Ref.~\cite{BordcochKozamehRojas2023}.  In this framework the radiative data
at null infinity are promoted to operators, while the spacetime geometry is
reconstructed from them through the null-surface equations.  The radiative
canonical algebra is the Ashtekar algebra at null infinity
\cite{Ashtekar1981,AshtekarStreubel1981}.  That work obtained the first
nontrivial quantum NSF scattering results and introduced
the scattering phase \(\delta T\) through \(S=\exp(\ii\delta T)\).

Refs.~\cite{QSI,QSII,QSIII} subsequently extended this construction by
resolving the graviton helicity sectors, completing the tree-level
scattering channels, and computing additional perturbative contributions,
including those that generate loop topologies.  In all these calculations
the outgoing operators are obtained directly from the quantum NSF field
equations rather than from a previously specified scattering operator.

This leads to the inverse problem studied here.  Given the perturbative
outgoing coefficients \(\delta a_n^{\rm out}\) supplied by quantum NSF, how
can one determine whether the resulting map is implemented by a formally
unitary operator \(S\)?  Preservation of the outgoing canonical commutation
relations is necessary, but it is not evident a priori that it is sufficient
to reconstruct a self-adjoint generator \(\delta T\).  Theorem~3 proves this
implication within the polynomial bosonic CCR algebra and provides an
explicit recursive construction of \(\delta T\) at every perturbative order.
The algebraic theorem is self-contained and independent of the calculations
in Refs.~\cite{QSI,QSII,QSIII}; only its NSF application uses the explicit
higher-order helicity kernels obtained there.
Ref.~\cite{QSII} verified the outgoing CCRs for the explicit NSF
coefficients through finite perturbative order.  The present work proves
the general converse reconstruction statement: within the polynomial
bosonic CCR algebra, preservation of the perturbative outgoing CCR
hierarchy is sufficient to reconstruct recursively a formal self-adjoint
BCH generator to all orders.

At order \(n\), the NSF cone integral defines a correction \(\delta a_n\),
homogeneous of degree \(n\) before operator contractions are taken into
account.  We first test whether the complete outgoing coefficients satisfy
the perturbative CCRs and then reconstruct the formal BCH generator.  The
NSF application also completes the direct order-three CCR test, including
its scalar sector, and reconstructs both the quartic and bilinear sectors of
\(\delta T_2\).  In particular, the canonical selection \(c=1\) and the
associated linear contraction are derived from the complete ordered NSF
coefficient rather than introduced as an independent correction.

Throughout this paper we use the flat solution-space specialization of
the NSF cone construction.  The variables \(x^a\) label the
four-dimensional solution space of the cut equation \(Z(x^a,\zeta,\bar
\zeta)\); they are not coordinates on a background spacetime supplied
independently of the NSF equations.  Spacetime geometry is reconstructed on
this solution space.  In the sector considered here, the reconstructed
metric is specialized to \(\eta_{ab}\); the radiative fields
are evaluated on the corresponding flat cuts \(Z_0\), and the associated
null generators, their geodesics, affine parameter, and cone measure are
kept at their flat-space values.  We refer to this choice below as the
fixed-Minkowski-cone or flat
solution-space specialization.  The auxiliary unitary operator acting on
the cut data is fixed to be
\begin{equation}
 \boxed{U=\mathrm{Id}.}
 \label{eq:introduction_U_identity}
\end{equation}
Consequently, we do not retain recursive-cut products involving
\(\delta Z_1\), nor perturbations of the metric or of the null geodesics.
The operator \(U\) in Eq.~\eqref{eq:introduction_U_identity} is not the
scattering operator \(S\) reconstructed below.

We adopt the Yang--Feldman working prescription used in
Ref.~\cite{QSIII,YangFeldman1950}.  The local NSF equations for
\(\Lambda\), \(\bar\Lambda\), \(\Omega\), and the reconstructed metric are
solved perturbatively with retarded boundary conditions, so that all local
nonlinear fields are functionals of the incoming radiative data.  The same
retarded solution is inserted in both the future and the
antipodally identified past cone representations.  The two cone terms
therefore do not define independent nonlinear solutions; they differ only
in the orientation of the outer affine integration.  At second order this
prescription is automatic because the source depends only on the free
incoming field, while at third order it means replacing every lower-order
field in the cubic source by its recursively determined incoming-data
expression.

The generic NSF kernels constructed in this way carry spatial-momentum
support but need not contain an independent four-dimensional
energy--momentum delta function.  The Poincar\'e/Weinberg scattering sector
is therefore a later kinematical reduction of the NSF map, rather than an
assumption built into the reconstruction theorem.

We reconstruct the scattering phase and use it directly in matrix
elements.  We write
\begin{equation}
 \boxed{
 S(\varepsilon)=\exp\!\bigl[\ii\delta T(\varepsilon)\bigr],
 \qquad
 \delta T(\varepsilon)^{\dagger}=\delta T(\varepsilon),
 }
 \label{eq:S_definition}
\end{equation}
through
\begin{equation}
 \boxed{
 a_{\lambda}^{\rm out}
 =S^{\dagger}a_{\lambda}^{\rm in}S .
 }
 \label{eq:unitary_implementation}
\end{equation}
All exponential and adjoint relations in this paper are understood as
formal power-series identities.  Once the coefficients of the
self-adjoint phase \(\delta T\) have been determined,
Eq.~\eqref{eq:unitary_implementation} generates the adjoint expansion and
preserves the canonical commutation relations through every perturbative
order reconstructed.  We do not address global Hilbert-space
implementability or the associated domain and closure properties.

The logical sequence is
\begin{equation}
 \boxed{
 \delta a_n
 \longrightarrow
 \delta a_n^{\rm out}
 \xrightarrow{\text{perturbative CCRs}}
 \delta T_{n-1}
 \longrightarrow
 S(\varepsilon)=e^{\ii\delta T(\varepsilon)}.
 }
 \label{eq:NSF_reconstruction_sequence}
\end{equation}
The central theorem states that complete coefficients satisfying the CCR
hierarchy determine a self-adjoint \(\delta T(\varepsilon)\) recursively.
The CCR test is performed directly on the known outgoing
coefficients and contains no unknown generator.  The later reconstruction
separates the new contribution at each order from the BCH iterations fixed
by lower orders.

We explicitly reconstruct the first two generator coefficients,
\(\delta T_1\) and \(\delta T_2\).  The quadratic NSF coefficient
determines the cubic generator \(\delta T_1\), while the
complete third-order coefficient, after subtraction of the lower-order BCH
iteration, determines both the quartic and bilinear sectors of
\(\delta T_2\).  The resulting formal exponential is then
used to organize the leading \(3\to1\) transition and connected
\(2\to2\) scattering into point contributions, canonical contractions
through on-shell intermediate states, and loop topologies.
Section~\ref{sec:adjoint_Duhamel} formulates the direct
adjoint map, the perturbative outgoing CCRs, and the inverse reconstruction
theorems.  Section~\ref{sec:recursive_inversion} introduces the NSF
coefficients and applies the CCR tests at orders two and three.
Sections~\ref{sec:deltaT1_reconstruction} and
\ref{sec:deltaT2_reconstruction} reconstruct the two generators, and the
subsequent sections apply them to scattering.  The explicit compatibility
checks and detailed matrix-element reductions are collected in the
appendices.

\section{CCR integrability and recursive BCH reconstruction}
\label{sec:adjoint_Duhamel}

This section establishes the central algebraic result used in the remainder
of the paper.  The input is a perturbative outgoing map
\(a_i\mapsto a_i^{\rm out}(\varepsilon)\), whose coefficients are assumed
known independently of any scattering generator.  We do not assume at the
outset that the map is generated by an operator \(S\).  Instead, we ask
whether preservation of the canonical commutation relations is sufficient
to reconstruct such a generator.

Theorem~3 identifies the perturbative CCR hierarchy, within the polynomial
realization of the bosonic CCR algebra, as the integrability condition for
the inverse adjoint problem.  After subtracting the BCH contribution fixed
at lower orders, comparison with the CCRs of the corresponding partial BCH
map forces the order-\(n\) residual to satisfy the homogeneous compatibility
conditions required for inverse-commutator reconstruction.  These
conditions determine the new coefficient \(\delta T_{n-1}\).

This result differs from the broader algebraic correspondence between
near-identity formal automorphisms and derivations,
and from the general theory of automorphisms of Weyl-type algebras
\cite{Praagman1986,BerestWilson2000,BagayokoEtAl2024}.  Here the starting
data are the perturbative outgoing coefficients themselves, and the
generator is recovered recursively from their CCR hierarchy.  The
reconstruction proceeds as follows:
\begin{equation}
\boxed{
\begin{aligned}
\{\delta a_n^{\rm out}\}
&\xrightarrow{\ {\rm BCH\ subtraction}\ }
\{R_n=\delta a_n^{\rm out}-I_n\}
\\
&\xrightarrow{\ {\rm outgoing\ CCRs}\ }
\{R_n\ {\rm satisfies\ the\ homogeneous\ integrability\ conditions}\}
\\
&\xrightarrow{\ {\rm Theorem~1}\ }
\{\delta T_{n-1}\}.
\end{aligned}
}
\label{eq:constructive_reconstruction_summary}
\end{equation}
Theorem~1 supplies the local inverse-commutator step, Theorem~2 gives the
order-by-order CCR equations obeyed by the known outgoing coefficients,
and Theorem~3 combines them into the all-order recursive reconstruction.

\subsection{Bosonic CCR algebra and the inverse adjoint problem}

We work in the normally ordered polynomial realization of the bosonic CCR
algebra generated by \(a(i)\) and \(a^\dagger(i)\).  This is the
creation--annihilation realization of the canonical commutation relations
used throughout bosonic Fock-space quantization
\cite{DerezinskiGerard2013,Gasiorowicz1966,Schwarz2024,LabuschagneMajewski2025}.  We use
the compact notation
\begin{equation}
i\equiv(\lambda_i,\vec k_i),
\qquad
a(i)\equiv a_{\lambda_i}(\vec k_i),
\qquad
a^\dagger(i)\equiv a_{\lambda_i}^\dagger(\vec k_i),
\label{eq:compact_leg_notation}
\end{equation}
and
\begin{equation}
\int_i\equiv\sum_{\lambda_i}\int d\mu_i,
\qquad
d\mu_i=\frac{d^3k_i}{2\omega_i},
\qquad
\omega_i=|\vec k_i|.
\label{eq:compact_integration_notation}
\end{equation}
The incoming operators satisfy
\begin{align}
[a(i),a^\dagger(j)]&=\Delta(i,j),
\label{eq:compact_CCR_deltaT1}
\\
[a(i),a(j)]&=0,
\qquad
[a^\dagger(i),a^\dagger(j)]=0,
\label{eq:compact_zero_CCR_deltaT1}
\end{align}
with the normalization
\begin{equation}
\int_i\Delta(j,i)F(i)=\frac12F(j).
\label{eq:half_contraction_rule}
\end{equation}

The complete outgoing map is written as
\begin{equation}
\boxed{
a_i^{\rm out}(\varepsilon)
=a_i+\sum_{n=2}^{\infty}
\varepsilon^{n-1}\delta a_{n,i}^{\rm out}.
}
\label{eq:deltaaout_plain_power}
\end{equation}
The inverse adjoint problem is to determine whether there exists a formal
self-adjoint series
\begin{equation}
\delta T(\varepsilon)
=\sum_{m=1}^{\infty}\varepsilon^m\delta T_m,
\qquad
\delta T_m^\dagger=\delta T_m,
\label{eq:deltaT_plain_power}
\end{equation}
such that
\begin{equation}
\boxed{
a^{\rm out}(\varepsilon)
=e^{-\ii\delta T(\varepsilon)}
a\,
e^{+\ii\delta T(\varepsilon)}
=e^{\ii\ad_{\delta T(\varepsilon)}}a,
\qquad
\ad_X(Y)\equiv[Y,X].
}
\label{eq:BCH_compact}
\end{equation}
The existence of \(\delta T\) is the conclusion to be reconstructed
from the outgoing coefficients, rather than an assumption used to obtain
them.

Expanding Eq.~\eqref{eq:BCH_compact} gives
\begin{equation}
a^{\rm out}-a
=\sum_{k=1}^{\infty}
\frac{\ii^k}{k!}\ad_{\delta T(\varepsilon)}^k a.
\label{eq:BCH_series}
\end{equation}
The first three nontrivial coefficients are
\begin{align}
\delta a_2^{\rm out}
&=\ii[a,\delta T_1],
\label{eq:BCH_order2_abstract}
\\
\delta a_3^{\rm out}
&=\ii[a,\delta T_2]
+\frac{\ii}{2}[\delta a_2^{\rm out},\delta T_1],
\label{eq:BCH_order3_abstract}
\\
\delta a_4^{\rm out}
&=\ii[a,\delta T_3]
+\frac{\ii}{2}
\left(
[\delta a_2^{\rm out},\delta T_2]
+[\delta a_3^{\rm out},\delta T_1]
\right)
+\frac{1}{12}
[[\delta a_2^{\rm out},\delta T_1],\delta T_1].
\label{eq:forward_commutator_first_orders}
\end{align}

At order \(n\), suppose that
\(\delta T_1,\ldots,\delta T_{n-2}\) have already been reconstructed.
Let \(I_{n,i}\) denote the coefficient at that order generated entirely by
these lower-order operators.  Thus
\begin{equation}
I_{2,i}=0,
\qquad
I_{3,i}=\frac{\ii}{2}
[\delta a_{2,i}^{\rm out},\delta T_1].
\label{eq:In_first_orders_abstract}
\end{equation}
The difference between the given outgoing coefficient and the BCH
contribution already forced by lower orders is
\begin{equation}
\boxed{
R_{n,i}\equiv\delta a_{n,i}^{\rm out}-I_{n,i}.
}
\label{eq:Rn_abstract_definition}
\end{equation}
Hence,
\begin{equation}
\delta a_{n,i}^{\rm out}
=
\underbrace{I_{n,i}}_{\text{fixed by lower orders}}
+
\underbrace{R_{n,i}}_{\text{new order-\(n\) information}}.
\label{eq:In_Rn_split}
\end{equation}
The inverse problem at order \(n\) reduces to
\begin{equation}
\boxed{
R_{n,i}=\ii[a_i,\delta T_{n-1}],
\qquad
\delta T_{n-1}^\dagger=\delta T_{n-1}.
}
\label{eq:Rn_inverse_commutator}
\end{equation}
If such a generator exists, Jacobi's identity and the centrality of the
incoming CCRs imply
\begin{align}
[R_{n,i},a_j]+[a_i,R_{n,j}]&=0,
\label{eq:Rn_aa_compatibility_sec2}
\\
[R_{n,i},a_j^\dagger]+[a_i,R_{n,j}^\dagger]&=0,
\label{eq:Rn_mixed_compatibility_sec2}
\end{align}
together with compatibility under Hermitian conjugation.  The first
ingredient of the reconstruction is that, within the polynomial Fock
algebra, these necessary conditions are also sufficient.

\subsection{Local integrability: the inverse commutator theorem}
\label{sec:inverse_commutator_solvability}

\paragraph{Theorem 1 (local inverse-commutator reconstruction).}
Let \(R_{[d]}(i)\) be a normally ordered homogeneous polynomial of operator
degree \(d\), with \(R_{[d]}(i)|0\rangle=0\), and let
\(\delta(a^\dagger(i))\) denote the corresponding correction of a creation
operator.  Within the polynomial Fock algebra, the following statements
are equivalent:
\begin{enumerate}
\item The corrections satisfy
\begin{align}
[R_{[d]}(i),a(j)]+[a(i),R_{[d]}(j)]&=0,
\label{eq:inverse_algebraic_aa}
\\
[R_{[d]}(i),a^\dagger(j)]
+[a(i),\delta(a^\dagger(j))]&=0,
\label{eq:inverse_algebraic_mixed}
\end{align}
and
\begin{equation}
\delta(a^\dagger(i))=R_{[d]}(i)^\dagger.
\label{eq:inverse_adjoint_compatibility}
\end{equation}
\item There exists a self-adjoint normally ordered operator
\(T_{[d]}\) such that
\begin{equation}
R_{[d]}(i)=\ii[a(i),T_{[d]}],
\qquad
R_{[d]}(i)^\dagger=\ii[a^\dagger(i),T_{[d]}].
\label{eq:inverse_problem_general}
\end{equation}
\end{enumerate}
The generator is unique up to an additive real multiple of the identity;
the normalization \(T_{[d]}|0\rangle=0\) fixes this ambiguity.

To construct it, decompose
\begin{equation}
R_{[d]}(i)=\sum_{r=0}^{d-1}R_{[d]}^{(r,d-r)}(i),
\label{eq:deltaan_sector_decomposition}
\end{equation}
where the superscript gives the number of creation and annihilation
operators.  Then
\begin{equation}
\boxed{
T_{[d]}
=-2\ii\sum_{r=0}^{d-1}\frac{1}{r+1}
\int_i a^\dagger(i)R_{[d]}^{(r,d-r)}(i).
}
\label{eq:all_order_generator_formula}
\end{equation}

\paragraph{Proof.}
If a self-adjoint \(T_{[d]}\) exists, the two commutator identities
follow immediately from Jacobi's identity and the centrality of the
incoming CCRs, while Hermitian conjugation gives the adjoint condition.
Conversely, in the sector with \(r\) creators, the first identity makes
the external leg and the \(r\) creator legs of the lifted monomial
completely symmetric.  The commutator with \(a(k)\) can therefore contract
any of the \(r+1\) creators in Eq.~\eqref{eq:all_order_generator_formula}.
Using Eq.~\eqref{eq:half_contraction_rule},
\begin{equation}
\ii\left[
a(k),
-\frac{2\ii}{r+1}\int_i
a^\dagger(i)R_{[d]}^{(r,d-r)}(i)
\right]
=R_{[d]}^{(r,d-r)}(k).
\label{eq:all_order_lifting_identity}
\end{equation}
Summing over \(r\) reconstructs \(R_{[d]}\).  The mixed identity and the
adjoint condition also imply self-adjointness explicitly.  Indeed, taking
the adjoint of the lifting identity gives
\(R_{[d]}(j)^\dagger=\ii[a^\dagger(j),T_{[d]}^\dagger]\).  Substitution of
this relation and \(R_{[d]}(i)=\ii[a(i),T_{[d]}]\) into
Eq.~\eqref{eq:inverse_algebraic_mixed}, followed by the Jacobi identity,
yields
\begin{equation}
[a(i),[a^\dagger(j),T_{[d]}^\dagger-T_{[d]}]]=0.
\label{eq:inverse_self_adjointness_mixed_derivative}
\end{equation}
For a normally ordered polynomial, this states that
\(T_{[d]}^\dagger-T_{[d]}\) can contain only a purely creating part, a
purely annihilating part, and a central constant.  Every monomial produced
by Eq.~\eqref{eq:all_order_generator_formula}, however, contains at least
one creator and one annihilator: the former is supplied by the explicit
\(a^\dagger(i)\), and the latter follows from
\(R_{[d]}(i)|0\rangle=0\).  The two pure sectors are therefore absent.
Moreover, both \(T_{[d]}\) and \(T_{[d]}^\dagger\) annihilate the vacuum,
so the remaining constant vanishes and
\(T_{[d]}^\dagger=T_{[d]}\).  Finally, two generators producing the same
residual differ by an operator commuting with every creator and
annihilator; irreducibility makes this difference a real multiple of the
identity. \hfill\(\square\)

Theorem~1 integrates the homogeneous commutator conditions into a normally
ordered generator.  This construction is related
to the broader literature on derivations of Weyl-type algebras
\cite{Sridharan1961,BenkartLopesOndrus2015} and to noncommutative
integrability questions formulated in terms of free and cyclic gradients
\cite{MaiSpeicher2019}; the explicit sector-by-sector lifting formula
\eqref{eq:all_order_generator_formula} is the form required here for the
perturbative CCR reconstruction.

If normal ordering produces several operator degrees at the same
perturbative order, the theorem is applied to each homogeneous component.
For
\[
R_n=\sum_d\sum_{r=0}^{d-1}R_{n,[d]}^{(r,d-r)},
\]
one obtains
\begin{equation}
\boxed{
\delta T_{n-1}
=-2\ii\sum_d\sum_{r=0}^{d-1}\frac{1}{r+1}
\int_i a^\dagger(i)R_{n,[d]}^{(r,d-r)}(i).
}
\label{eq:all_degree_generator_formula}
\end{equation}
Equivalently, at perturbative order \(n\) the theorem is applied to every
component \(R_{n,[d]}\); the sum of the corresponding local lifts is the
coefficient \(\delta T_{n-1}=\sum_d T_{n,[d]}\).  Thus \(n\) labels
perturbative order throughout, whereas \(d\) labels operator degree.
This extension is needed below because the normally ordered order-three
coefficient has operator degrees three and one.

It remains to prove that the residual \(R_n\) defined by
Eq.~\eqref{eq:Rn_abstract_definition} satisfies the homogeneous conditions
of Theorem~1.  This follows from the outgoing CCR hierarchy.

\subsection{The perturbative outgoing CCR hierarchy}
\label{sec:perturbative_CCR_reconstruction}
%=====================================================================

The complete outgoing coefficients are known before any BCH generator is
reconstructed.  Their canonical consistency can therefore be tested
directly, without introducing the unknown operators \(\delta T_m\).  Write
\begin{equation}
a_i^{\rm out}(\varepsilon)
=a_i+\sum_{n=2}^{\infty}
\varepsilon^{n-1}\delta a_{n,i}^{\rm out},
\qquad
a_i^{{\rm out}\dagger}(\varepsilon)
=a_i^\dagger+\sum_{n=2}^{\infty}
\varepsilon^{n-1}\delta a_{n,i}^{{\rm out}\dagger}.
\label{eq:theorem2_out_expansion}
\end{equation}

\paragraph{Theorem 2 (perturbative CCR hierarchy).}
The outgoing operators in Eq.~\eqref{eq:theorem2_out_expansion} satisfy
\begin{equation}
[a_i^{\rm out},a_j^{\rm out}]=0,
\qquad
[a_i^{\rm out},a_j^{{\rm out}\dagger}]=\Delta(i,j),
\qquad
[a_i^{{\rm out}\dagger},a_j^{{\rm out}\dagger}]=0,
\label{eq:theorem2_exact_outgoing_CCRs}
\end{equation}
if and only if their perturbative coefficients satisfy, for every
coefficient \(\delta a_n^{\rm out}\) (order \(\varepsilon^{n-1}\)),
\begin{align}
0={}&
[\delta a_{n,i}^{\rm out},a_j]
+[a_i,\delta a_{n,j}^{\rm out}]
+\sum_{r=2}^{n-1}
[\delta a_{r,i}^{\rm out},
 \delta a_{n+1-r,j}^{\rm out}],
\label{eq:theorem2_CCR_aa_general}
\\
0={}&
[\delta a_{n,i}^{\rm out},a_j^\dagger]
+[a_i,\delta a_{n,j}^{{\rm out}\dagger}]
+\sum_{r=2}^{n-1}
[\delta a_{r,i}^{\rm out},
 \delta a_{n+1-r,j}^{{\rm out}\dagger}],
\label{eq:theorem2_CCR_mixed_general}
\end{align}
together with the adjoint of
Eq.~\eqref{eq:theorem2_CCR_aa_general}.

At order two these equations reduce to
\begin{align}
[\delta a_{2,i}^{\rm out},a_j]
+[a_i,\delta a_{2,j}^{\rm out}]&=0,
\label{eq:theorem2_order2_aa}
\\
[\delta a_{2,i}^{\rm out},a_j^\dagger]
+[a_i,\delta a_{2,j}^{{\rm out}\dagger}]&=0.
\label{eq:theorem2_order2_mixed}
\end{align}
At order three they become
\begin{align}
[\delta a_{3,i}^{\rm out},a_j]
+[a_i,\delta a_{3,j}^{\rm out}]
+[\delta a_{2,i}^{\rm out},\delta a_{2,j}^{\rm out}]&=0,
\label{eq:theorem2_order3_aa}
\\
[\delta a_{3,i}^{\rm out},a_j^\dagger]
+[a_i,\delta a_{3,j}^{{\rm out}\dagger}]
+[\delta a_{2,i}^{\rm out},
  \delta a_{2,j}^{{\rm out}\dagger}]&=0.
\label{eq:theorem2_order3_mixed}
\end{align}
At order four, the quadratic term in each equation is replaced by
the two cross terms
\([\delta a_2,\delta a_3]+[\delta a_3,\delta a_2]\), with the
corresponding dagger on the second entry in the mixed sector.

\paragraph{Proof.}
Substituting Eq.~\eqref{eq:theorem2_out_expansion} into
Eq.~\eqref{eq:theorem2_exact_outgoing_CCRs} and equating each coefficient
of \(\varepsilon^{n-1}\) to zero gives
Eqs.~\eqref{eq:theorem2_CCR_aa_general} and
\eqref{eq:theorem2_CCR_mixed_general}.  Conversely, if this hierarchy and
its adjoint hold at every order, every nonconstant coefficient in the
three outgoing commutators vanishes, and
Eq.~\eqref{eq:theorem2_exact_outgoing_CCRs} follows as a formal power
series. \hfill\(\square\)

Theorem~2 is independent of the inverse reconstruction and tests only the
known outgoing coefficients.  Its role in the main theorem is that the
inhomogeneous terms in
Eqs.~\eqref{eq:theorem2_CCR_aa_general} and
\eqref{eq:theorem2_CCR_mixed_general} depend exclusively on lower-order
coefficients.  Once those lower orders have been reconstructed by BCH,
the partial BCH map and the given outgoing map have exactly the same
inhomogeneous CCR source at the next order.  Their difference therefore
obeys the homogeneous conditions of Theorem~1.

\subsection{Main result: CCR-to-generator reconstruction}

Formal exponential--logarithm correspondences between automorphisms and
derivations provide an abstract route from a near-identity algebra
automorphism to an infinitesimal generator in suitable completed settings
\cite{Praagman1986,BagayokoEtAl2024}.  Theorem~3 gives a coefficient-level
statement adapted to the present bosonic CCR problem:
the perturbative CCR equations themselves supply, order by order, the
homogeneous integrability conditions needed to reconstruct each new BCH
generator coefficient.

\paragraph{Theorem 3 (CCR integrability and recursive BCH reconstruction).}
Suppose that the complete outgoing coefficients in
Eq.~\eqref{eq:theorem2_out_expansion} satisfy the perturbative CCR hierarchy
of Theorem~2, including the adjoint compatibility conditions.  Suppose
also that, after normal ordering and subtraction of the BCH terms fixed at
lower orders, the residuals belong to the polynomial Fock algebra covered
by Theorem~1.

Then the outgoing map admits a recursively reconstructible formal
self-adjoint generator
\begin{equation}
\delta T(\varepsilon)
=\sum_{m=1}^{\infty}\varepsilon^m\delta T_m,
\qquad
\delta T_m^\dagger=\delta T_m,
\label{eq:theorem3_generator_series}
\end{equation}
unique at each order up to an additive real multiple of the identity,
fixed by the vacuum normalization, such that
\begin{equation}
\boxed{
a^{\rm out}(\varepsilon)
=\exp\!\left[-\ii\delta T(\varepsilon)\right]
a\,
\exp\!\left[+\ii\delta T(\varepsilon)\right].
}
\label{eq:theorem2_unitary_reconstruction}
\end{equation}
Once
\(\delta T_1,\ldots,\delta T_{n-2}\) have been reconstructed, their BCH
expansion determines \(I_n\).  The residual
\begin{equation}
R_{n,i}
=\delta a_{n,i}^{\rm out}-I_{n,i}
\label{eq:theorem3_recursive_residual}
\end{equation}
automatically satisfies the homogeneous compatibility conditions of
Theorem~1 and hence
\begin{equation}
R_{n,i}=\ii[a_i,\delta T_{n-1}].
\label{eq:theorem3_recursive_commutator}
\end{equation}
The new generator is obtained constructively from
\begin{equation}
\boxed{
\delta T_{n-1}
=-2\ii\sum_d\sum_{r=0}^{d-1}\frac{1}{r+1}
\int_i a^\dagger(i)R_{n,[d]}^{(r,d-r)}(i).
}
\label{eq:theorem3_recursive_generator}
\end{equation}

\paragraph{Proof.}
At order two the sums in
Eqs.~\eqref{eq:theorem2_CCR_aa_general} and
\eqref{eq:theorem2_CCR_mixed_general} are empty.  Since
\(I_2=0\), one has
\begin{equation}
R_2=\delta a_2^{\rm out}.
\end{equation}
The order-two CCR equations are therefore precisely the homogeneous
compatibility conditions of Theorem~1.  It follows that there exists a
self-adjoint \(\delta T_1\) such that
\begin{equation}
\delta a_{2,i}^{\rm out}=\ii[a_i,\delta T_1].
\label{eq:theorem2_base_T1}
\end{equation}
This establishes the base step.

The first nontrivial instance of the recursive mechanism occurs at
order three.  The adjoint map generated only by the already
reconstructed \(\delta T_1\) is
\begin{equation}
\widetilde a_i(\varepsilon)
=e^{-\ii\varepsilon\delta T_1}a_i
 e^{+\ii\varepsilon\delta T_1}
=a_i+\varepsilon\delta a_{2,i}^{\rm out}
+\varepsilon^2 I_{3,i}+\mathcal O(\varepsilon^3),
\qquad
I_{3,i}=\frac{\ii}{2}
[\delta a_{2,i}^{\rm out},\delta T_1].
\label{eq:theorem2_auxiliary_order3}
\end{equation}
Because this is an exact adjoint conjugation, it preserves the incoming
CCRs:
\begin{equation}
[\widetilde a_i(\varepsilon),\widetilde a_j(\varepsilon)]
=e^{-\ii\varepsilon\delta T_1}[a_i,a_j]
 e^{+\ii\varepsilon\delta T_1}=0.
\label{eq:theorem2_auxiliary_CCR_exact}
\end{equation}
The coefficient of \(\varepsilon^2\) gives
\begin{equation}
[I_{3,i},a_j]+[a_i,I_{3,j}]
+[\delta a_{2,i}^{\rm out},\delta a_{2,j}^{\rm out}]=0.
\label{eq:theorem2_I3_CCR_aa}
\end{equation}
The same coefficient in the CCR of the given outgoing map is
\begin{equation}
[\delta a_{3,i}^{\rm out},a_j]
+[a_i,\delta a_{3,j}^{\rm out}]
+[\delta a_{2,i}^{\rm out},\delta a_{2,j}^{\rm out}]=0.
\label{eq:theorem2_deltaa3_CCR_aa}
\end{equation}
The inhomogeneous lower-order source is identical in the two equations.
Subtracting them therefore removes it and leaves
\begin{equation}
[R_{3,i},a_j]+[a_i,R_{3,j}]=0,
\qquad
R_{3,i}\equiv\delta a_{3,i}^{\rm out}-I_{3,i}.
\label{eq:theorem2_R3_homogeneous_aa}
\end{equation}
Likewise, comparing
\([\widetilde a_i,\widetilde a_j^\dagger]=\Delta(i,j)\) with the mixed
outgoing CCR gives
\begin{equation}
[R_{3,i},a_j^\dagger]+[a_i,R_{3,j}^\dagger]=0.
\label{eq:theorem2_R3_homogeneous_mixed}
\end{equation}
Thus \(R_3\) satisfies exactly the hypotheses of Theorem~1, which
reconstructs a self-adjoint \(\delta T_2\) such that
\[
R_{3,i}=\ii[a_i,\delta T_2].
\]
This calculation exhibits the mechanism that persists at subsequent
orders: the CCR hierarchy cancels the lower-order BCH source and converts
the new residual into a homogeneous inverse-commutator problem.

For the induction step, assume that
\(\delta T_1,\ldots,\delta T_{n-2}\) have already been reconstructed and
define the partial generator
\begin{equation}
\delta T_{<n}(\varepsilon)
=\sum_{m=1}^{n-2}\varepsilon^m\delta T_m
\end{equation}
and the corresponding partial adjoint map
\begin{equation}
S_{<n}(\varepsilon)
=\exp\!\left[\ii\delta T_{<n}(\varepsilon)\right].
\label{eq:theorem2_partial_unitary}
\end{equation}
By the induction hypothesis its adjoint action agrees with the given
outgoing map through order \(n-1\), while at the next order it
produces a definite BCH coefficient \(I_n\):
\begin{equation}
S_{<n}^\dagger a_iS_{<n}
=a_i+\sum_{r=2}^{n-1}
\varepsilon^{r-1}\delta a_{r,i}^{\rm out}
+\varepsilon^{n-1}I_{n,i}
+\mathcal O(\varepsilon^n).
\label{eq:theorem2_partial_expansion}
\end{equation}
The partial adjoint map preserves the CCR identically.  Since it agrees
with the given outgoing map at all lower orders, the order-\(n\) CCR
equations for \(I_n\) contain exactly the same inhomogeneous sums as the
order-\(n\) CCR equations for \(\delta a_n^{\rm out}\).  Subtraction
therefore gives
\begin{equation}
R_{n,i}\equiv\delta a_{n,i}^{\rm out}-I_{n,i},
\qquad
[R_{n,i},a_j]+[a_i,R_{n,j}]=0,
\qquad
[R_{n,i},a_j^\dagger]+[a_i,R_{n,j}^\dagger]=0.
\label{eq:theorem2_Rn_homogeneous}
\end{equation}
Together with the adjoint condition, these are precisely the hypotheses of
Theorem~1.  Hence the new generator coefficient exists and is reconstructed
explicitly by Eq.~\eqref{eq:theorem3_recursive_generator}:
\begin{equation}
R_{n,i}=\ii[a_i,\delta T_{n-1}].
\end{equation}
Adding \(\varepsilon^{n-1}\delta T_{n-1}\) to the generator extends the
agreement between the BCH map and the given outgoing map by one order.
This completes the induction and proves the reconstruction to all
perturbative orders. \hfill\(\square\)

At order four, for example,
\begin{equation}
\begin{aligned}
S_{<4}
&=e^{\ii(\varepsilon\delta T_1+
\varepsilon^2\delta T_2)}
\\
&\Longrightarrow\quad
R_4\equiv\delta a_4^{\rm out}-I_4
\text{ satisfies the homogeneous CCR conditions}
\\
&\Longrightarrow\quad
R_4=\ii[a,\delta T_3].
\end{aligned}
\label{eq:theorem2_n4_one_line}
\end{equation}
No new integrability condition has to be imposed at this order beyond the
outgoing CCR hierarchy itself.

\subsection{Interpretation and scope of the reconstruction theorem}

Within the stated polynomial Fock-algebra hypotheses, the perturbative CCRs
both test whether the independently constructed outgoing coefficients define
a canonical map and suffice to reconstruct the formal adjoint generator
order by order.  The construction is summarized in
Eq.~\eqref{eq:constructive_reconstruction_summary}.

The theorem also separates the information carried by each perturbative
order.  The term \(I_n\) is fixed completely by the generators already
determined at lower orders and therefore contains only the BCH iteration
of previously reconstructed interactions.  The residual \(R_n\) is the
part not fixed by those iterations, and Eq.~\eqref{eq:theorem3_recursive_generator}
converts it into the new generator coefficient
\(\delta T_{n-1}\).  The inverse problem can therefore be solved without
guessing the form of the scattering phase and without deriving the
outgoing coefficients from a preassigned \(S\)-operator.

The statement is algebraic and perturbative.  In particular,
Eq.~\eqref{eq:theorem2_unitary_reconstruction} establishes a formal
self-adjoint BCH generator within the polynomial Fock algebra assumed in
Theorem~1.  Questions concerning domains, closures, and unitary
implementability on a completed infinite-dimensional Fock Hilbert space
are analytically distinct from the formal algebraic reconstruction; this
distinction is standard in the mathematical theory of CCR representations
and becomes essential in field-theoretic canonical transformations
\cite{DerezinskiGerard2013,TorreVaradarajan1999}.  These analytic questions
are not required for the perturbative reconstruction developed here.

The NSF application proceeds in two stages.  First, the NSF construction
supplies the coefficients \(\delta a_n^{\rm out}\), and their outgoing CCRs are tested
directly.  Once those tests hold, Theorem~3 reconstructs the corresponding
\(\delta T_{n-1}\).  In this way the dynamical calculation determines the
outgoing map, while the CCR hierarchy provides the integrability mechanism
that recovers its BCH generator.

\section{NSF coefficients and their perturbative CCR test}
\label{sec:recursive_inversion}

Section~\ref{sec:adjoint_Duhamel} treated the coefficients
\(\delta a_n^{\rm out}\) as abstract input data.  We now introduce the
coefficients supplied by the NSF field equations and ask whether they
satisfy the perturbative outgoing CCRs.

The NSF hierarchy is obtained by solving the vacuum Einstein equations
\(R_{ab}=0\) recursively, as developed in Refs.~\cite{QSI,QSII,QSIII}.
We denote the radiative coefficient at order \(n\) by
\begin{equation}
\boxed{
\delta a_{n,\lambda}(q)
\equiv
-\Pi^{\rm rad}_{\lambda,q}
\left[
 Z_{n,{\rm cone}}^{\rm out}
 +\mathcal A Z_{n,{\rm cone}}^{\rm in}
\right].
}
\label{eq:NSF_cone_kernel_definition}
\end{equation}
Here \(Z_{n,{\rm cone}}\) is the order-\(n\) cone contribution.  In both
terms of Eq.~\eqref{eq:NSF_cone_kernel_definition}, the local nonlinear
source is evaluated on the same retarded Yang--Feldman solution determined
by the incoming radiative data.  The superscripts ``out'' and ``in'' label
the two cone representations and their opposite outer affine
prescriptions; they do not denote two independently reconstructed nonlinear
fields.  The classical amplitudes are replaced by incoming oscillators with
the chosen ordering, after which the complete expression is put in normal
form.  Hence
\(\delta a_n\) includes every lower operator-degree term generated by
that ordering.

At order two, the outgoing coefficient is
\begin{equation}
\delta a_2^{\rm out}=\delta a_2.
\label{eq:NSF_order2_outgoing}
\end{equation}
Its canonical test is therefore
\begin{align}
[\delta a_{2,i},a_j]+[a_i,\delta a_{2,j}]&=0,
\label{eq:NSF_order2_CCR_aa}
\\
[\delta a_{2,i},a_j^\dagger]
+[a_i,\delta a_{2,j}^\dagger]&=0.
\label{eq:NSF_order2_CCR_mixed}
\end{align}
These equations were verified for the quadratic NSF kernel.  By
Theorem~3, their validity guarantees the existence of a self-adjoint
\(\delta T_1\); its explicit reconstruction is deferred to the next
section.

At order three, the sector-by-sector calculation in
Appendix~\ref{app:compatibility_conditions} gives
\begin{equation}
\delta a_3^{\rm out}=\delta a_3.
\label{eq:NSF_order3_outgoing_general}
\end{equation}
The complete coefficient must satisfy
\begin{align}
0={}&
[\delta a_{3,i}^{\rm out},a_j]
+[a_i,\delta a_{3,j}^{\rm out}]
+[\delta a_{2,i},\delta a_{2,j}],
\label{eq:NSF_order3_CCR_aa}
\\
0={}&
[\delta a_{3,i}^{\rm out},a_j^\dagger]
+[a_i,\delta a_{3,j}^{{\rm out}\dagger}]
+[\delta a_{2,i},\delta a_{2,j}^\dagger].
\label{eq:NSF_order3_CCR_mixed}
\end{align}
These are the order-three CCR equations.  Their left-hand sides
are evaluated directly with the complete ordered NSF coefficients and no
unknown generator.  The verification includes the
operator-degree-zero condition and all independent degree-two sectors.

The remaining analysis first evaluates the CCR equations for the NSF
coefficients and then uses Theorems~1 and~3 to reconstruct the self-adjoint
generators of their formal unitary implementation.

%=====================================================================
\section{Reconstruction of the first generator
\texorpdfstring{$\delta T_1$}{delta T1}}
\label{sec:deltaT1_reconstruction}
%=====================================================================

At the first nonlinear order there is no lower-order BCH iteration.  The
quadratic coefficient in the outgoing map is therefore the NSF coefficient
itself,
\begin{equation}
a^{\mathrm{out}}(i)
=
a(i)+\varepsilon\,\delta a_2(i)
+\mathcal O(\varepsilon^2),
\qquad
\delta a_2(i)=\ii[a(i),\delta T_1].
\label{eq:first_order_asymptotic_map}
\end{equation}
After frequency separation, normal ordering, and relabeling of the
difference channel, the positive-frequency NSF correction takes the form
\begin{equation}
\boxed{
\begin{aligned}
\delta a_2(3)
={}&
\int_1\int_2
\mathcal K^{aa}(3;1,2)\,
a(1)a(2)
\\
&+
\int_1\int_2
\mathcal K^{ca}(3,1|2)\,
a^\dagger(1)a(2).
\end{aligned}
}
\label{eq:deltaa2_general_compact}
\end{equation}
The fixed label \(3\) is not integrated.  The complete kernels include the
spatial momentum distributions, mode normalizations, helicity projectors,
angular Green functions, principal-value factors, and Bose
symmetrizations inherited from the NSF construction.  Their respective
supports are
\begin{align}
\mathcal K^{aa}(3;1,2)
&\propto
\delta^{(3)}
\!\left(\vec k_3-\vec k_1-\vec k_2\right),
\label{eq:Kaa_spatial_support_prop}
\\
\mathcal K^{ca}(3,1|2)
&\propto
\delta^{(3)}
\!\left(\vec k_3+\vec k_1-\vec k_2\right).
\label{eq:Kca_spatial_support_prop}
\end{align}
Only the symmetric part of the two-annihilator kernel contributes, so that
\begin{equation}
\mathcal K^{aa}(3;1,2)=\mathcal K^{aa}(3;2,1).
\label{eq:Kaa_internal_symmetry_prop}
\end{equation}

Appendix~\ref{app:compatibility_conditions} verifies directly that the
complete distributional kernels in Eq.~\eqref{eq:deltaa2_general_compact}
satisfy the order-two CCR conditions of Theorem~2.  Applying the lifting
formula of Theorem~1 separately to the two operator sectors gives
\begin{equation}
\boxed{
\begin{aligned}
\delta T_1
={}&-2\ii\int_{312}
K(3;1,2)a^\dagger(3)a(1)a(2)
\\
&-\ii\int_{312}
L(3,1|2)a^\dagger(3)a^\dagger(1)a(2),
\end{aligned}
}
\label{eq:NSF_T1_all_sectors}
\end{equation}
where \(K=\mathcal K^{aa}\) and \(L=\mathcal K^{ca}\).  Using the two
order-two identities in Eq.~\eqref{eq:app_E2_kernel_relations}, the same
operator can be written in the manifestly self-adjoint form
\begin{equation}
\boxed{
\delta T_1
=-\ii\int_{312}
L(3,1|2)a^\dagger(3)a^\dagger(1)a(2)
+\mathrm{H.c.}
}
\label{eq:NSF_T1_final_Kca}
\end{equation}
Indeed, the Hermitian-conjugate term in
Eq.~\eqref{eq:NSF_T1_final_Kca} is precisely the first line of
Eq.~\eqref{eq:NSF_T1_all_sectors}; it reproduces the \(aa\) sector of
Eq.~\eqref{eq:deltaa2_general_compact} and is not an additional
interaction.  Both cubic monomials contain at least one annihilator, so
the vacuum normalization is automatic.  Consequently,
\begin{equation}
\boxed{
\delta a_2(4)=\ii[a(4),\delta T_1],
\qquad
\delta T_1^\dagger=\delta T_1,
\qquad
\delta T_1|0\rangle=0.
}
\label{eq:NSF_deltaa2_from_T1_corollary}
\end{equation}
Thus the quadratic NSF kernel determines the complete three-graviton
vertex entering the first term of the scattering generator.

%=====================================================================
\section{Reconstruction of the second generator
\texorpdfstring{$\delta T_2$}{delta T2}}
\label{sec:deltaT2_reconstruction}
%=====================================================================

At order three, the NSF input is the quantized radiative
coefficient \(\delta a_3\) defined by
Eq.~\eqref{eq:NSF_cone_kernel_definition}.  Its normally ordered cubic
part is supplemented by the linear contraction generated by the ordering
of the mixed cubic monomial.  The classical kernel \(B\) fixes the
coefficient of this monomial but does not select the placement of its one
creator among its two annihilators.  Quantization therefore leaves the
three possibilities \(c=0,1,2\).  Appendix~\ref{app:compatibility_conditions}
keeps the three possible placements explicitly through a coefficient
\(c=0,1,2\) and proves, without using the order-three mixed CCR as an
assumption, that
\begin{equation}
E^{a\bar a;(0)}_{3,ij}(c)
=
\frac{1-c}{2}
\int_{rs}K(i;r,s)K(j;r,s)^*.
\label{eq:main_scalar_ordering_selection}
\end{equation}
Equation~\eqref{eq:app_E3_mixed_degree0} fixes \(c=1\) uniquely for the
nontrivial NSF kernel; it is not an external ordering prescription.  At
this value the induced linear
contraction cancels the degree-zero double-contraction contribution from
\([\delta a_2,\delta a_2^\dagger]\).  Thus the CCR, rather than the
classical expression alone, selects the canonical ordering.  The linear
kernel \(D_1\) is derived from \(B\), and the cancellation follows from the
order-two kernel identities.  Together with the already
verified normally ordered degree-two sectors, the resulting coefficient
satisfies the complete outgoing CCRs through this order.  No operator
beyond this ordered coefficient is added:
\begin{equation}
\boxed{\delta a_3^{\rm out}=\delta a_3\big|_{c=1}.}
\label{eq:deltaa3_out_equals_NSF}
\end{equation}

The order-three BCH equation is not
\(\delta a_3=\ii[a,\delta T_2]\), because the iteration generated by
\(\delta T_1\) is already present.  Define
\begin{equation}
\boxed{
I_3\equiv\frac{\ii}{2}[\delta a_2,\delta T_1],
\qquad
R_3\equiv\delta a_3-I_3.
}
\label{eq:deltaT2_NSF_equation}
\end{equation}
Theorem~3 guarantees that \(R_3\) satisfies the homogeneous reconstruction
conditions, and the second generator is determined by
\begin{equation}
\boxed{R_3=\ii[a,\delta T_2].}
\label{eq:deltaT2_residual_equation}
\end{equation}
Here \(R_3\) is only the BCH-subtracted coefficient used in the inverse
step; it is not a correction of the NSF result.

For the explicit normal-ordering calculation, introduce
\begin{equation}
K(q;u,v)\equiv\mathcal K^{aa}(q;u,v),
\qquad
L(q,u|v)\equiv\mathcal K^{ca}(q,u|v).
\label{eq:KL_abbreviations_I3}
\end{equation}
The quadratic compatibility relations imply
\begin{equation}
K(q;u,v)=K(q;v,u),
\qquad
L(q,u|v)=L(u,q|v),
\qquad
K(q;u,v)=-\frac12L(u,v|q)^*.
\label{eq:KL_relations_I3}
\end{equation}
Normal ordering the commutator in
Eq.~\eqref{eq:deltaT2_NSF_equation} gives
\begin{equation}
I_3^{(3)}
=I_3^{(0,3)}+I_3^{(1,2)}+I_3^{(2,1)},
\label{eq:I3_cubic_sector_split}
\end{equation}
with
\begin{align}
I_3^{(0,3)}(q)
={}&-\frac12\int_{uvrs}
K(q;u,v)L(r,s|v)^*a(u)a(r)a(s),
\label{eq:I3_aaa_kernel}
\\
I_3^{(1,2)}(q)
={}&\int_{uvst}
K(q;u,v)L(v,s|t)a^\dagger(s)a(u)a(t)
\nonumber\\
&-\frac14\int_{uvrs}
L(q,u|v)L(r,s|v)^*a^\dagger(u)a(r)a(s)
\nonumber\\
&+\frac12\int_{uvrt}
L(q,u|v)L(u,r|t)^*a^\dagger(t)a(r)a(v),
\label{eq:I3_caa_kernel}
\\
I_3^{(2,1)}(q)
={}&\frac12\int_{uvst}
L(q,u|v)L(v,s|t)a^\dagger(u)a^\dagger(s)a(t)
\nonumber\\
&-\frac14\int_{uvrs}
L(q,u|v)L(r,s|u)a^\dagger(r)a^\dagger(s)a(v).
\label{eq:I3_cca_kernel}
\end{align}
The double contraction is the linear term
\begin{equation}
\boxed{
I_3^{(1)}(q)
=\frac14\int_{uvt}K(q;u,v)L(u,v|t)a(t)
=-\frac18\int_{uvt}L(u,v|q)^*L(u,v|t)a(t).
}
\label{eq:I3_linear_explicit}
\end{equation}
It is useful to denote its one-operator kernel by
\begin{equation}
\boxed{
D_I(q|t)
\equiv
-\frac18\int_{uv}L(u,v|q)^*L(u,v|t),
\qquad
I_3^{(1)}(q)=\int_tD_I(q|t)a(t).
}
\label{eq:DI_kernel_explicit}
\end{equation}
There is no \(a^\dagger\) term in \(I_3^{(1)}\).

The complete normal-ordered NSF coefficient can be written as
\begin{equation}
\boxed{
\begin{aligned}
\delta a_3(4)
={}&
\int_1\int_2\int_3
(2\pi)^3\delta^{(3)}
\!\left(\vec k_4-\vec k_1-\vec k_2-\vec k_3\right)
\left[\mathcal A(4;1,2,3)\right]_{(123)}a(1)a(2)a(3)
\\
&+
\int_1\int_2\int_3
(2\pi)^3\delta^{(3)}
\!\left(\vec k_4+\vec k_1-\vec k_2-\vec k_3\right)
\left[\mathcal B(4,1|2,3)\right]_{(23)}
a^\dagger(1)a(2)a(3)
\\
&+
\int_1\int_2\int_3
(2\pi)^3\delta^{(3)}
\!\left(\vec k_4+\vec k_1+\vec k_2-\vec k_3\right)
\left[\mathcal C(4,1,2|3)\right]_{(12)}
a^\dagger(1)a^\dagger(2)a(3)
\\
&+
\int_r D(4|r)a(r).
\end{aligned}
}
\label{eq:deltaa3_ABC_decomposition}
\end{equation}
The three kernels contain the common mode normalization
\begin{equation}
\mathcal N^{(3)}(4;1,2,3)
=
\frac{2\omega_4}{\nu(\omega_4)}
\nu(\omega_1)\nu(\omega_2)\nu(\omega_3),
\qquad
\nu(\omega)=8\pi^2\sqrt{4\pi G}\,\omega^{-3/2},
\label{eq:N3_complete_for_BCH_test}
\end{equation}
together with the helicity projectors, angular Green functions, and
principal-value factors.  Parenthesized subscripts denote normalized
symmetrization over the indicated labels; thus \(\mathcal A\) is
symmetric in \((1,2,3)\), \(\mathcal B\) in \((2,3)\), and
\(\mathcal C\) in \((1,2)\).  The chosen vacuum-preserving ordering has no
term proportional to \(a^\dagger(r)\), as stated explicitly in
Eq.~\eqref{eq:app_deltaa3_D}.

Let \(\mathcal I_A,\mathcal I_B,\mathcal I_C\) denote the complete
distributional kernels obtained from
Eqs.~\eqref{eq:I3_aaa_kernel}--\eqref{eq:I3_cca_kernel} after the same
normalized Bose symmetrizations as in
Eq.~\eqref{eq:deltaa3_ABC_decomposition}.  Define the BCH-subtracted
kernels
\begin{equation}
\boxed{
\mathcal A_R=\mathcal A-\mathcal I_A,\qquad
\mathcal B_R=\mathcal B-\mathcal I_B,\qquad
\mathcal C_R=\mathcal C-\mathcal I_C,\qquad
H=D-D_I.
}
\label{eq:R3_residual_ABC_kernels}
\end{equation}
Thus
\begin{equation}
\begin{aligned}
R_3(q)
={}&\int_{123}\mathcal A_R(q;1,2,3)a(1)a(2)a(3)
\\
&+\int_{123}\mathcal B_R(q,1|2,3)
a^\dagger(1)a(2)a(3)
\\
&+\int_{123}\mathcal C_R(q,1,2|3)
a^\dagger(1)a^\dagger(2)a(3)
+\int_rH(q|r)a(r).
\end{aligned}
\label{eq:R3_complete_operator_decomposition}
\end{equation}
All kernels in this equation include their momentum distributions,
normalizations, and helicity labels.

Applying Eq.~\eqref{eq:all_order_generator_formula} to every operator
sector gives the fully lifted second generator:
\begin{equation}
\boxed{
\begin{aligned}
\delta T_2
={}&-2\ii\int_{q123}
\mathcal A_R(q;1,2,3)
a^\dagger(q)a(1)a(2)a(3)
\\
&-\ii\int_{q123}
\mathcal B_R(q,1|2,3)
a^\dagger(q)a^\dagger(1)a(2)a(3)
\\
&-\frac{2\ii}{3}\int_{q123}
\mathcal C_R(q,1,2|3)
a^\dagger(q)a^\dagger(1)a^\dagger(2)a(3)
\\
&-2\ii\int_{qr}H(q|r)a^\dagger(q)a(r).
\end{aligned}
}
\label{eq:deltaT2_general_kernel_form}
\end{equation}
The factors \(2\), \(1\), \(2/3\), and \(2\) are fixed by the number of
creator legs that may contract with the external annihilator and by the
half-contraction normalization in Eq.~\eqref{eq:half_contraction_rule}.

\paragraph{The four-leg sector.}
The first three lines of
Eq.~\eqref{eq:deltaT2_general_kernel_form} are the complete quartic lift
of the cubic part of \(R_3\).  The quantities
\(\mathcal I_A,\mathcal I_B,\mathcal I_C\) are not new dynamical kernels:
they are given explicitly by
Eqs.~\eqref{eq:I3_aaa_kernel}--\eqref{eq:I3_cca_kernel} and are fixed
entirely by \(K\) and \(L\).  Thus the four-leg part of \(\delta T_2\)
is obtained by the three definite subtractions in
Eq.~\eqref{eq:R3_residual_ABC_kernels}.  The order-three CCR calculation
in Appendix~\ref{app:compatibility_conditions}, together with
Theorem~3, guarantees that these lifted sectors combine into a
self-adjoint quartic operator.  No additional kernel is introduced.

\paragraph{The quadratic sector.}
The scalar CCR sector is not needed to reconstruct the quartic vertex
\(\delta T_2^{(4)}\), which is fixed by the cubic residual.  It is,
however, required for compatibility of the complete third-order outgoing
coefficient and for self-adjointness of the bilinear sector
\(\delta T_2^{(2)}\) associated with the linear residual.
The scalar CCR in Appendix~\ref{app:compatibility_conditions} gives
\begin{equation}
D(i|j)+D(j|i)^*
=-\int_{rs}K(i;r,s)K(j;r,s)^*.
\label{eq:D_scalar_CCR_repeated}
\end{equation}
On the other hand, Eqs.~\eqref{eq:DI_kernel_explicit} and
\eqref{eq:KL_relations_I3} imply
\begin{equation}
D_I(i|j)+D_I(j|i)^*
=-\int_{rs}K(i;r,s)K(j;r,s)^*.
\label{eq:DI_scalar_identity}
\end{equation}
Subtracting the two equations yields
\begin{equation}
\boxed{H(i|j)+H(j|i)^*=0.}
\label{eq:H_antihermitian}
\end{equation}
Hence the last line of Eq.~\eqref{eq:deltaT2_general_kernel_form} is
self-adjoint.  Explicitly, the quadratic part is
\begin{equation}
\boxed{
\delta T_2^{(2)}
=-2\ii\int_{qr}
\left[
D(q|r)
+\frac18\int_{uv}L(u,v|q)^*L(u,v|r)
\right]
a^\dagger(q)a(r).
}
\label{eq:deltaT2_final_quadratic}
\end{equation}
The complete result is
\begin{equation}
\boxed{\delta T_2=\delta T_2^{(4)}+\delta T_2^{(2)},}
\label{eq:deltaT2_quartic_plus_quadratic}
\end{equation}
where \(\delta T_2^{(4)}\) is the sum of the first three lines of
Eq.~\eqref{eq:deltaT2_general_kernel_form}.  Every monomial is normally
ordered and contains at least one annihilator on the right, so
\begin{equation}
\boxed{
\delta T_2^\dagger=\delta T_2,
\qquad
\delta T_2|0\rangle=0.
}
\label{eq:deltaT2_hermitian_vacuum}
\end{equation}
No \(aa+a^\dagger a^\dagger\) term is allowed: hermiticity would require
the pair-creation term, which does not annihilate the vacuum.  Likewise,
an \(aa^\dagger\) monomial must first be normal ordered into
\(a^\dagger a\) plus a scalar, and the scalar is excluded by the vacuum
normalization.  Thus \(a^\dagger a\) is the only possible quadratic
sector.  If a particular ordering gives \(D=D_I\), this quadratic sector
vanishes; otherwise Eq.~\eqref{eq:deltaT2_final_quadratic} is the
one-particle canonical rotation required to reproduce the complete
one-operator part of \(\delta a_3\).

\section{The \texorpdfstring{$3\to1$}{3 to 1} transition}
\label{subsec:main_three_to_one_selection}

The distinction between the $\delta T_2$ term and the lower-order BCH
iteration has a direct diagrammatic interpretation in the $3\to1$ matrix
element worked out in Appendix~\ref{app:three_to_one_selection}.
Figure~\ref{fig:three_to_one_T2_T1_squared} shows the
contribution generated by $\ii\delta T_2$ and the three tree contributions
generated by canonical contractions in $-\tfrac12\delta T_1^2$.

\begin{figure}[t]
\centering
\begin{tikzpicture}[
  ext/.style={decorate,decoration={snake,amplitude=1.0pt,segment length=5pt},
    line width=0.7pt},
  int/.style={decorate,decoration={snake,amplitude=1.15pt,segment length=5pt},
    line width=1.0pt},
  vert/.style={circle,fill=black,inner sep=2.2pt},
  every node/.style={font=\small}
]

% Contribution from delta T2
\begin{scope}[shift={(0,3.4)}]
  \node[vert] (v) at (0,0) {};
  \draw[ext] (-2.2,1.05) -- node[above,sloped] {$k_1,\lambda_1$} (v);
  \draw[ext] (-2.4,0) -- node[above] {$k_2,\lambda_2$} (v);
  \draw[ext] (-2.2,-1.05) -- node[below,sloped] {$k_3,\lambda_3$} (v);
  \draw[ext] (v) -- node[above] {$k_4,\lambda_4$} (2.5,0);
  \node at (0,-1.45) {(a) Contribution from $\ii\delta T_2$};
\end{scope}

% Contracted channel 123
\begin{scope}[shift={(-5.1,0)},scale=0.76,transform shape]
  \node[vert] (l) at (0,0) {};
  \node[vert] (r) at (2.5,0) {};
  \draw[ext] (-1.8,0.9) -- node[above,sloped] {$k_2,\lambda_2$} (l);
  \draw[ext] (-1.8,-0.9) -- node[below,sloped] {$k_3,\lambda_3$} (l);
  \draw[int] (l) -- node[above] {$r_{23},\lambda_{r_{23}}$} (r);
  \draw[ext] (1.1,-1.45) -- node[below,sloped] {$k_1,\lambda_1$} (r);
  \draw[ext] (r) -- node[above] {$k_4,\lambda_4$} (4.25,0);
  \node at (1.2,-2.0) {(b) $123$: $\lambda_2=-\lambda_3$};
\end{scope}

% Contracted channel 231
\begin{scope}[shift={(0,0)},scale=0.76,transform shape]
  \node[vert] (l) at (0,0) {};
  \node[vert] (r) at (2.5,0) {};
  \draw[ext] (-1.8,0.9) -- node[above,sloped] {$k_3,\lambda_3$} (l);
  \draw[ext] (-1.8,-0.9) -- node[below,sloped] {$k_1,\lambda_1$} (l);
  \draw[int] (l) -- node[above] {$r_{31},\lambda_{r_{31}}$} (r);
  \draw[ext] (1.1,-1.45) -- node[below,sloped] {$k_2,\lambda_2$} (r);
  \draw[ext] (r) -- node[above] {$k_4,\lambda_4$} (4.25,0);
  \node at (1.2,-2.0) {(c) $231$: $\lambda_3=-\lambda_1$};
\end{scope}

% Contracted channel 312
\begin{scope}[shift={(5.1,0)},scale=0.76,transform shape]
  \node[vert] (l) at (0,0) {};
  \node[vert] (r) at (2.5,0) {};
  \draw[ext] (-1.8,0.9) -- node[above,sloped] {$k_1,\lambda_1$} (l);
  \draw[ext] (-1.8,-0.9) -- node[below,sloped] {$k_2,\lambda_2$} (l);
  \draw[int] (l) -- node[above] {$r_{12},\lambda_{r_{12}}$} (r);
  \draw[ext] (1.1,-1.45) -- node[below,sloped] {$k_3,\lambda_3$} (r);
  \draw[ext] (r) -- node[above] {$k_4,\lambda_4$} (4.25,0);
  \node at (1.2,-2.0) {(d) $312$: $\lambda_1=-\lambda_2$};
\end{scope}
\end{tikzpicture}
\caption{Feynman-like tree diagrams for the $3\to1$ matrix
element.  Panel (a) is the four-leg vertex reconstructed from
$\delta T_2$.  Panels (b)--(d) are the three BCH contributions obtained by
canonical contractions in $\delta T_1^2$.  The left legs are incoming and
$k_4$ is outgoing.  Each $r_{ij}$ is on shell and carries the contraction factor
$1/(2\omega_{ij})$, not an off-shell covariant propagator.}
\label{fig:three_to_one_T2_T1_squared}
\end{figure}
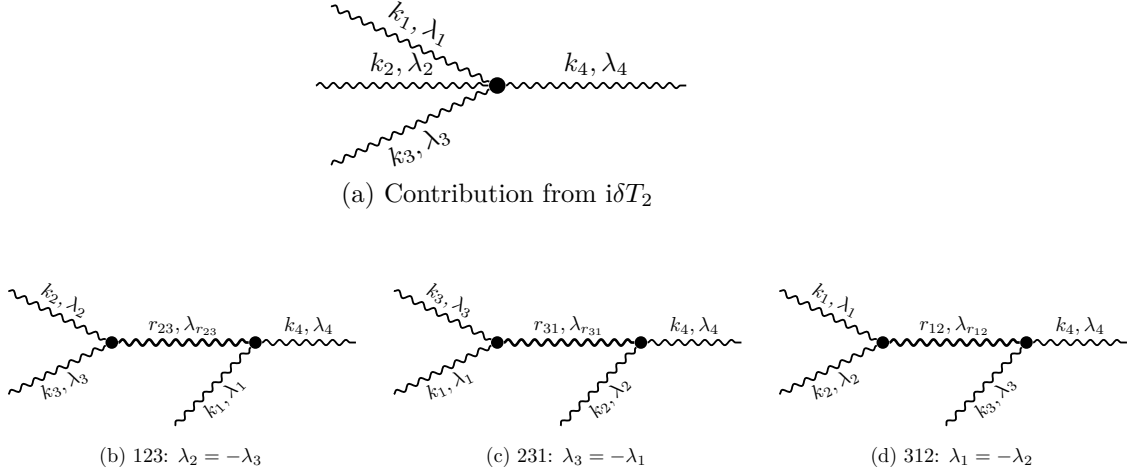

For completeness, we record the helicity-resolved amplitudes derived in
Appendix~\ref{app:three_to_one_selection}.  Their point term is defined by
the ordered residual kernel \(\mathcal A_R\).  Their common spatial-momentum
support is
\begin{equation}
 \vec k_4=\vec k_1+\vec k_2+\vec k_3,
 \qquad
 \omega_4=|\vec k_1+\vec k_2+\vec k_3|,
 \label{eq:main_three_to_one_support}
\end{equation}
with no additional delta function imposing four-dimensional energy
conservation.  For equal incoming helicities, the three contracted tree contributions
vanish and the explicit evaluation of the point kernel gives
\(\mathcal A_{R,\lambda_4;+++}=\mathcal A_{R,\lambda_4;---}=0\).
Therefore the complete equal-helicity amplitudes vanish:
\begin{equation}
\boxed{
 \mathscr M^{(2)}_{\lambda_4;+++}=0,
 \qquad
 \mathscr M^{(2)}_{\lambda_4;---}=0,
 \qquad
 \lambda_4=\pm.
}
\label{eq:main_three_to_one_equal_helicity_zero}
\end{equation}

The mixed-helicity configurations are not excluded.  With
\begin{equation}
 r_{ij}=\bigl(|\vec k_i+\vec k_j|,\vec k_i+\vec k_j\bigr),
 \qquad
 \omega_{ij}=|\vec k_i+\vec k_j|,
 \label{eq:main_rij_definition}
\end{equation}
the two independent amplitudes are
\begin{equation}
\boxed{
\begin{aligned}
&\mathscr M^{(2)}_{\lambda_4;++-}(4;1,2,3)
\\
&\quad=
\delta^{(3)}
(\vec k_4-\vec k_1-\vec k_2-\vec k_3)
\Bigg\{
\frac34
\alpha_{\lambda_4;++-}(4;1,2,3)
\\
&\qquad\quad+
\frac14
\frac{
\kappa^{aa}_{\lambda_4;+-}(4;1,r_{23})
\kappa^{aa}_{-;+-}(r_{23};2,3)
}{2\omega_{23}}
\\
&\qquad\quad+
\frac14
\frac{
\kappa^{aa}_{\lambda_4;+-}(4;2,r_{31})
\kappa^{aa}_{-;-+}(r_{31};3,1)
}{2\omega_{31}}
\Bigg\},
\qquad
\lambda_4=\pm,
\end{aligned}
}
\label{eq:main_three_to_one_mixed_plus}
\end{equation}
and
\begin{equation}
\boxed{
\begin{aligned}
&\mathscr M^{(2)}_{\lambda_4;--+}(4;1,2,3)
\\
&\quad=
\delta^{(3)}
(\vec k_4-\vec k_1-\vec k_2-\vec k_3)
\Bigg\{
\frac34
\alpha_{\lambda_4;--+}(4;1,2,3)
\\
&\qquad\quad+
\frac14
\frac{
\kappa^{aa}_{\lambda_4;-+}(4;1,r_{23})
\kappa^{aa}_{+;-+}(r_{23};2,3)
}{2\omega_{23}}
\\
&\qquad\quad+
\frac14
\frac{
\kappa^{aa}_{\lambda_4;-+}(4;2,r_{31})
\kappa^{aa}_{+;+-}(r_{31};3,1)
}{2\omega_{31}}
\Bigg\},
\qquad
\lambda_4=\pm.
\end{aligned}
}
\label{eq:main_three_to_one_mixed_minus}
\end{equation}
The other four Bose-related orderings are
\begin{align}
\mathscr M^{(2)}_{\lambda_4;+-+}(4;1,2,3)
&=
\mathscr M^{(2)}_{\lambda_4;++-}(4;1,3,2),
&
\mathscr M^{(2)}_{\lambda_4;-++}(4;1,2,3)
&=
\mathscr M^{(2)}_{\lambda_4;++-}(4;2,3,1),
\nonumber\\
\mathscr M^{(2)}_{\lambda_4;-+-}(4;1,2,3)
&=
\mathscr M^{(2)}_{\lambda_4;--+}(4;1,3,2),
&
\mathscr M^{(2)}_{\lambda_4;+--}(4;1,2,3)
&=
\mathscr M^{(2)}_{\lambda_4;--+}(4;2,3,1).
\label{eq:main_three_to_one_mixed_permutations}
\end{align}
Equations~\eqref{eq:main_three_to_one_mixed_plus}--
\eqref{eq:main_three_to_one_mixed_permutations} give the complete
mixed-helicity kernels.  These sectors are not eliminated by the helicity
selection rule, although their value at a particular kinematic point is
fixed by the corresponding component of \(\mathcal A_R\).

%=====================================================================
\section{The \texorpdfstring{$2\to2$}{2 to 2} scattering process}
\label{sec:two_to_two_scattering}
%=====================================================================

\begin{figure}[t]
\centering
\begin{tikzpicture}[
  scale=0.625,
  ext/.style={decorate,decoration={snake,amplitude=1.0pt,segment length=5pt},
    line width=0.75pt},
  yarn/.style={line width=0.65pt},
  every node/.style={font=\small}
]
  \draw[fill=black!7,draw=black,line width=0.8pt] (0,0) circle (1.18);
  \begin{scope}
    \clip (0,0) circle (1.08);
    \draw[yarn] (-1.05,0.10) .. controls (-0.55,0.95) and (0.45,-0.95) .. (1.02,-0.10);
    \draw[yarn] (-0.95,-0.55) .. controls (-0.20,0.25) and (0.20,0.95) .. (0.92,0.45);
    \draw[yarn] (-0.82,0.68) .. controls (-0.05,-0.20) and (0.35,-0.75) .. (0.94,-0.62);
    \draw[yarn] (-0.45,-1.00) .. controls (0.70,-0.20) and (-0.65,0.30) .. (0.52,1.00);
    \draw[yarn] (-1.00,-0.18) .. controls (-0.05,-0.72) and (0.62,0.72) .. (1.00,0.16);
  \end{scope}
  \draw[ext] (-3.2,1.05) -- node[above,sloped] {$k_1,\lambda_1$} (-1.02,0.48);
  \draw[ext] (-3.2,-1.05) -- node[below,sloped] {$k_2,\lambda_2$} (-1.02,-0.48);
  \draw[ext] (1.02,0.48) -- node[above,sloped] {$k_3,\lambda_3$} (3.2,1.05);
  \draw[ext] (1.02,-0.48) -- node[below,sloped] {$k_4,\lambda_4$} (3.2,-1.05);
  \node[fill=white,inner sep=1.5pt] at (0,0)
  {$\mathscr M^{\rm conn}_{2\to2}$};
\end{tikzpicture}
\caption{The complete connected $2\to2$ on-shell kernel before its
resolution into vertices and canonical contractions.  The interior blob
denotes the sum of all connected terms generated by the reconstructed
scattering operator; it is not an additional interaction vertex.}
\label{fig:two_to_two_complete_blob}
\end{figure}
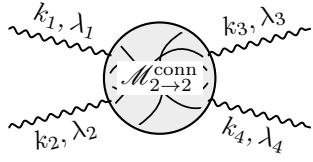

The next application is the connected four-leg matrix element.  Before
resolving it into individual contractions, it is useful to represent the
complete object by the four-leg on-shell kernel shown in
Fig.~\ref{fig:two_to_two_complete_blob}.

\subsection{Diagrammatic rules}
\label{subsec:diagrammatic_rules}

The reconstructed generators define on-shell operator rules analogous to
Feynman rules \cite{Gasiorowicz1966}, but without introducing off-shell
covariant propagators.
They may be summarized as follows.
\begin{enumerate}
\item \emph{Vertices.}  The generator $\delta T_{n-1}$ supplies an
$(n+1)$-graviton vertex.  Each normally ordered sector specifies the
assignment of incoming and outgoing legs, and its momentum-helicity kernel
is the vertex kernel.

\item \emph{External legs.}  An annihilation operator represents an incoming
external graviton and a creation operator an outgoing external graviton.
All physical external energies are positive, $\omega_i=|\vec k_i|$.

\item \emph{Vertex factors and multiplicities.}  A single insertion carries
the factor $\ii\delta T_{n-1}$.  Products of vertices and their coefficients
are obtained by expanding $S=\exp(\ii\sum_m\delta T_m)$.  Distinct operator
orderings are retained until the contractions have been evaluated.

\item \emph{Internal lines.}  An internal line is a canonical contraction.
It carries an on-shell momentum, a sum over the intermediate helicity, and
the contraction factor $1/(2\omega_{\vec q})$ in the normalization used here.

\item \emph{Momentum support and channels.}  The spatial delta functions in
the vertex kernels determine the internal momenta and overall support.  The
$s$, $t$, and $u$ channels are the three inequivalent pairings of four
external legs.  No four-dimensional energy delta function is added unless
it follows from the NSF matrix element being considered.

\item \emph{Crossing.}  Moving a graviton between the initial and final
states interchanges its creation and annihilation operators.  In signed
momentum notation,
\begin{equation}
\boxed{
(\vec k,\lambda)_{\rm out}
\longleftrightarrow
(-\vec k,-\lambda)_{\rm in},
\qquad \omega=|\vec k|.
}
\label{eq:NSF_external_crossing_rule}
\end{equation}
The helicity changes sign because it is defined relative to the physical
direction of propagation.

\item \emph{Connected diagrams and loops.}  Only connected contractions
contribute to the connected amplitude.  Vacuum diagrams are removed by the
vacuum normalization.  For a connected graph with $I$ internal lines and
$V$ vertices, $L=I-V+1$.
\end{enumerate}

Here and below, ``loop'' denotes the topology generated by NSF on-shell
canonical contractions, counted by \(L=I-V+1\).  This terminology does not identify the
resulting kernels with conventional off-shell Feynman loop integrals and,
by itself, does not imply ultraviolet finiteness.  Establishing such a
comparison requires the Poincar\'e restriction together with a separate
analysis of the NSF measures and kernels; this is distinct from the usual
off-shell perturbative-gravity discussion
\cite{tHooftVeltman1974,GoroffSagnotti1986,Donoghue1994}.

Crossing one incoming leg of the $3\to1$ kernel converts its
$a^\dagger aaa$ sector into the $a^\dagger a^\dagger aa$ sector required
for $2\to2$ scattering.  Appendix~\ref{app:two_to_two_crossing} verifies
this statement directly in the explicit NSF kernels.

The figures below give a direct resolution of the blob in
Fig.~\ref{fig:two_to_two_complete_blob}: the point term comes from
$\ii\delta T_2$, the three tree channels come from
$-\tfrac12\delta T_1^2$, and the loop diagrams come from higher powers in
the same exponential.  Their formulas are given below together with the
corresponding diagrams; no additional vertex is being defined.

\subsection{Direct \texorpdfstring{$\delta a_3\,\delta a_3$}{delta a3 delta a3}
sectors selected by the canonical ordering}
\label{subsec:direct_deltaa3_deltaa3_sectors}

The order-three outgoing coefficient entering the direct four-graviton
matrix element must be the same canonical operator that satisfies the
Ashtekar CCR.  The scalar identity proved in
Appendix~\ref{app:cubic_compatibility} selects the symmetric ordering
\(c=1\), for which
\begin{equation}
\boxed{
\delta a_{3,i}
=
\delta a_{3,i}^{(3)}
+
\delta a_{3,i}^{(1)},
\qquad
\delta a_{3,i}^{(1)}
=
\int_r D_1(i|r)a(r),
\qquad
D_1(i|r)=\frac12\int_x B(i,x|x,r).
}
\label{eq:main_deltaa3_canonical_split}
\end{equation}
Here the superscripts \((3)\) and \((1)\) denote operator degree after
normal ordering, not perturbative order.  The kernel \(B\) is the mixed
cubic kernel multiplying \(a^\dagger aa\), and the last equality is the
linear self-contraction generated by the symmetric ordering.

For two incoming gravitons \(1,2\) and two outgoing labels \(3,4\), define
the direct \((33)\) contribution by
\begin{equation}
\boxed{
\mathscr M_{33}(3,4;1,2)
=
\langle0|
\delta a_{3}(3)\,
\delta a_{3}(4)\,
a^\dagger(1)a^\dagger(2)
|0\rangle .
}
\label{eq:main_M33_full_canonical_definition}
\end{equation}
Substituting Eq.~\eqref{eq:main_deltaa3_canonical_split} before making any
connectedness restriction gives four operator-degree sectors,
\begin{equation}
\boxed{
\mathscr M_{33}
=
\mathscr M_{33}^{(3,3)}
+\mathscr M_{33}^{(3,1)}
+\mathscr M_{33}^{(1,3)}
+\mathscr M_{33}^{(1,1)} .
}
\label{eq:main_M33_four_degree_sectors}
\end{equation}
These terms are distinct from the perturbative labels on
\(\delta a_n\): both factors in every term are still the third-order
outgoing coefficient.

The new contractions can be evaluated without introducing any additional
vertex.  With the normalization used throughout this paper,
\begin{equation}
\int_u\Delta(u,r)F(u)=\frac12F(r).
\label{eq:main_half_contraction_recalled_M33}
\end{equation}
For the displayed Wightman ordering in
Eq.~\eqref{eq:main_M33_full_canonical_definition}, the nonzero mixed
operator-degree sector has the linear factor on the left and the mixed
cubic \(B\) branch on the right:
\begin{align}
\mathscr M_{33}^{(1,3)}(3,4;1,2)
={}&
\int_{urst}
D_1(3|u)B(4,r|s,t)
\nonumber\\
&\times
\langle0|
a(u)a^\dagger(r)a(s)a(t)
a^\dagger(1)a^\dagger(2)
|0\rangle .
\label{eq:main_M13_before_contractions}
\end{align}
The first elementary contraction gives
\[
a(u)a^\dagger(r)
=
a^\dagger(r)a(u)+\Delta(u,r),
\]
and the term beginning with \(a^\dagger(r)\) annihilates the left vacuum.
The remaining two annihilators contract with the two incoming creators:
\begin{align}
&\langle0|
a(u)a^\dagger(r)a(s)a(t)
a^\dagger(1)a^\dagger(2)
|0\rangle
\nonumber\\
&\qquad=
\Delta(u,r)
\left[
\Delta(s,1)\Delta(t,2)
+\Delta(s,2)\Delta(t,1)
\right].
\label{eq:main_M13_explicit_contractions}
\end{align}
Using Eq.~\eqref{eq:main_half_contraction_recalled_M33} three times and
the Bose symmetry \(B(4,r|s,t)=B(4,r|t,s)\) yields
\begin{equation}
\boxed{
\mathscr M_{33}^{(1,3)}(3,4;1,2)
=
\frac14\int_r
D_1(3|r)\,B(4,r|1,2)
=
\frac18\int_{rx}
B(3,x|x,r)\,B(4,r|1,2).
}
\label{eq:main_M13_reduced}
\end{equation}
The exchange \(3\leftrightarrow4\) supplies the partner attachment of the
two outgoing gravitons.

In the opposite operator order one has
\begin{align}
\mathscr M_{33}^{(3,1)}(3,4;1,2)
={}&
\int_{rstu}
B(3,r|s,t)D_1(4|u)
\nonumber\\
&\times
\langle0|
a^\dagger(r)a(s)a(t)a(u)
a^\dagger(1)a^\dagger(2)
|0\rangle
=0 .
\label{eq:main_M31_wightman_zero}
\end{align}
Indeed, acting from the right, the three annihilators
\(a(s)a(t)a(u)\) act on a two-particle state before the creator
\(a^\dagger(r)\) can act.  Equation~\eqref{eq:main_M31_wightman_zero}
refers to the displayed Wightman order; exchanging the two outgoing
insertions gives the corresponding nonzero partner already represented by
Eq.~\eqref{eq:main_M13_reduced} with \(3\leftrightarrow4\).

Finally, the linear--linear sector is
\begin{align}
\mathscr M_{33}^{(1,1)}(3,4;1,2)
={}&
\int_{rs}D_1(3|r)D_1(4|s)
\nonumber\\
&\times
\langle0|a(r)a(s)a^\dagger(1)a^\dagger(2)|0\rangle
\nonumber\\
={}&
\boxed{
\frac14\left[
D_1(3|1)D_1(4|2)
+
D_1(3|2)D_1(4|1)
\right]} .
\label{eq:main_M11_reduced}
\end{align}
This contribution factorizes into two one-particle corrections and
therefore drops from the connected four-graviton amplitude, although it is
part of the full matrix element generated by the canonical outgoing map.

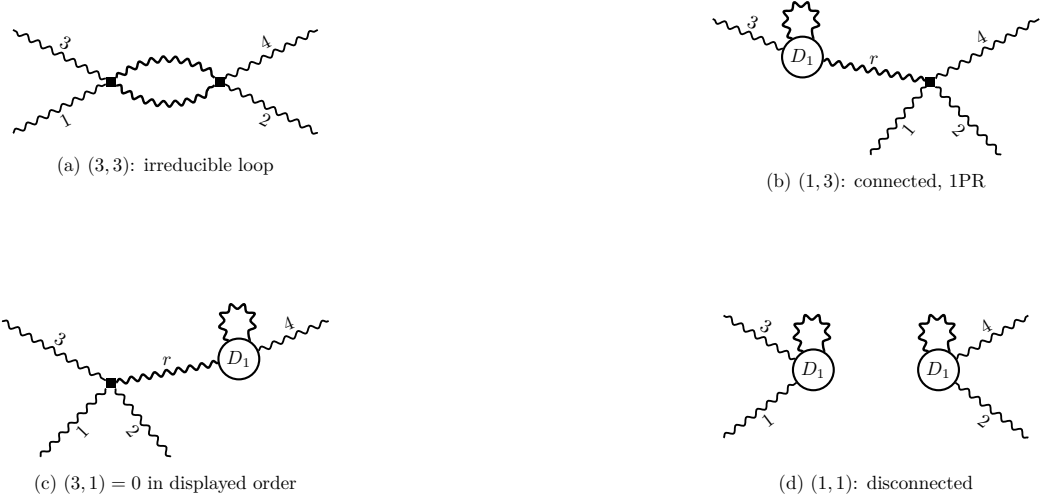
\begin{figure}[t]
\centering
\begin{tikzpicture}[
  ext/.style={decorate,decoration={snake,amplitude=1.0pt,segment length=5pt},
    line width=0.7pt},
  int/.style={decorate,decoration={snake,amplitude=1.05pt,segment length=4.5pt},
    line width=0.9pt},
  Bvert/.style={rectangle,fill=black,inner sep=2.5pt},
  Dvert/.style={circle,draw=black,fill=white,inner sep=2.2pt,line width=0.8pt},
  every node/.style={font=\small}
]

% (3,3)
\begin{scope}[shift={(-4.7,3.0)},scale=0.72,transform shape]
  \node[Bvert] (l) at (-1.0,0) {};
  \node[Bvert] (r) at (1.0,0) {};
  \draw[ext] (-2.8,-0.95) -- node[below,sloped] {$1$} (l);
  \draw[ext] (l) -- node[above,sloped] {$3$} (-2.8,0.95);
  \draw[int] (l) to[bend left=36] (r);
  \draw[int] (l) to[bend right=36] (r);
  \draw[ext] (2.8,-0.95) -- node[below,sloped] {$2$} (r);
  \draw[ext] (r) -- node[above,sloped] {$4$} (2.8,0.95);
  \node at (0,-1.55) {(a) $(3,3)$: irreducible loop};
\end{scope}

% (1,3)
\begin{scope}[shift={(4.7,3.0)},scale=0.72,transform shape]
  \node[Dvert] (d) at (-1.35,0.45) {$D_1$};
  \node[Bvert] (b) at (1.0,0) {};
  \draw[ext] (d) -- node[above,sloped] {$3$} (-3.0,1.15);
  \draw[int] (d) -- node[above] {$r$} (b);
  \draw[ext] (-0.1,-1.35) -- node[below,sloped] {$1$} (b);
  \draw[ext] (2.3,-1.35) -- node[below,sloped] {$2$} (b);
  \draw[ext] (b) -- node[above,sloped] {$4$} (3.0,1.15);
  \draw[int] (d) to[out=120,in=60,looseness=6] (d);
  \node at (0,-1.85) {(b) $(1,3)$: connected, 1PR};
\end{scope}

% (3,1)
\begin{scope}[shift={(-4.7,-1.0)},scale=0.72,transform shape]
  \node[Bvert] (b) at (-1.0,0) {};
  \node[Dvert] (d) at (1.35,0.45) {$D_1$};
  \draw[ext] (-3.0,1.15) -- node[above,sloped] {$3$} (b);
  \draw[ext] (-2.3,-1.35) -- node[below,sloped] {$1$} (b);
  \draw[ext] (0.1,-1.35) -- node[below,sloped] {$2$} (b);
  \draw[int] (b) -- node[above] {$r$} (d);
  \draw[ext] (d) -- node[above,sloped] {$4$} (3.0,1.15);
  \draw[int] (d) to[out=120,in=60,looseness=6] (d);
  \node at (0,-1.85) {(c) $(3,1)=0$ in displayed order};
\end{scope}

% (1,1)
\begin{scope}[shift={(4.7,-1.0)},scale=0.72,transform shape]
  \node[Dvert] (d1) at (-1.15,0.25) {$D_1$};
  \node[Dvert] (d2) at (1.15,0.25) {$D_1$};
  \draw[ext] (-2.8,-1.0) -- node[below,sloped] {$1$} (d1);
  \draw[ext] (d1) -- node[above,sloped] {$3$} (-2.8,1.25);
  \draw[ext] (2.8,-1.0) -- node[below,sloped] {$2$} (d2);
  \draw[ext] (d2) -- node[above,sloped] {$4$} (2.8,1.25);
  \draw[int] (d1) to[out=120,in=60,looseness=6] (d1);
  \draw[int] (d2) to[out=120,in=60,looseness=6] (d2);
  \node at (0,-1.85) {(d) $(1,1)$: disconnected};
\end{scope}
\end{tikzpicture}
\caption{Direct \(2\to2\) sectors generated by the product of two canonical
third-order outgoing coefficients.  Panel (a) is the cubic--cubic
irreducible one-loop topology.  Panel (b) is the new linear--cubic
contribution; when \(D_1=\tfrac12\int B\) is expanded, the small loop at the
\(D_1\) insertion represents the ordering self-contraction, while the line
joining \(D_1\) to the cubic kernel makes the graph connected but
one-particle reducible.  Panel (c) vanishes in the displayed Wightman
ordering; the exchanged outgoing ordering gives the partner of panel (b).
Panel (d) is the factorized product of two one-particle corrections and is
absent from the connected amplitude.}
\label{fig:canonical_deltaa3_deltaa3_sectors}
\end{figure}

Figure~\ref{fig:canonical_deltaa3_deltaa3_sectors} displays all four
operator-degree sectors.  In the diagrams a filled square denotes the
mixed cubic \(B\) kernel with its outgoing NSF label included, whereas a
filled circle labelled \(D_1\) denotes the two-leg insertion generated by
the symmetric self-contraction.  Expanding the circle according to
\(D_1(i|r)=\tfrac12\int_xB(i,x|x,r)\) reveals a tadpole-like internal
contraction at that insertion.  Thus the mixed term is connected but
one-particle reducible; the \((1,1)\) term is disconnected and factorized.
The \((3,3)\) term contains the irreducible one-loop topology already
present when only the normally ordered cubic sector is retained.

The complete canonical \((33)\) matrix element and its connected part are
therefore distinguished:
\begin{align}
\mathscr M_{33}^{\rm full}
={}&
\mathscr M_{33}^{(3,3)}
+
\mathscr M_{33}^{(1,3)}
+
(3\leftrightarrow4)
+
\mathscr M_{33}^{(1,1)},
\label{eq:main_M33_full_with_linear}
\\
\mathscr M_{33}^{\rm conn}
={}&
\mathscr M_{33}^{(3,3)}
+
\mathscr M_{33}^{(1,3)}
+
(3\leftrightarrow4),
\label{eq:main_M33_connected_with_linear}
\end{align}
where the second line still contains the one-particle-reducible
linear--cubic terms.  Restricting further to the irreducible four-leg
kernel leaves only \(\mathscr M_{33}^{(3,3)}\).  This separation explains
why the linear contraction is indispensable for the canonical
\(a^{\rm in}\mapsto a^{\rm out}\) map and its CCR, while it does not alter
the irreducible cubic--cubic loop kernel.

\subsection{Tree and loop contributions}
\label{subsec:two_to_two_scattering}

The two reconstructed generators define the formal unitary series
\begin{equation}
\boxed{
S^{[2]}
=
\exp\!\left[\ii\left(\delta T_1+\delta T_2\right)\right].
}
\label{eq:S_two_reconstructed_generators}
\end{equation}
The label $[2]$ means that the NSF generators have been reconstructed
through $\delta T_2$; the exponential itself is not truncated.
The lowest nontrivial connected four-leg matrix element is
\begin{equation}
\boxed{
\begin{aligned}
\mathscr M^{(2)}_{\lambda_3\lambda_4;\lambda_1\lambda_2}
={}&
\left\langle k_3,\lambda_3;k_4,\lambda_4\right|
\left(-\frac12\delta T_1^2+\ii\delta T_2\right)
\\
&\hspace{37mm}\times
\left|k_1,\lambda_1;k_2,\lambda_2\right\rangle_{\rm conn}.
\end{aligned}
}
\label{eq:two_to_two_degree_two_matrix_element}
\end{equation}

The four-leg part of the second generator is
\begin{equation}
\boxed{
\begin{aligned}
\left.\ii\delta T_2\right|_{a^{\dagger2}a^2}
={}&
\int_1\int_2\int_3\int_4
(2\pi)^3\delta^{(3)}
\!\left(\vec k_1+\vec k_2-\vec k_3-\vec k_4\right)
\\
&\times
\left[\mathcal B_R(3,4|1,2)\right]_{(34)(12)}
a^\dagger(3)a^\dagger(4)a(1)a(2).
\end{aligned}
}
\label{eq:two_to_two_iT2_sector}
\end{equation}
There are $2!\,2!=4$ contractions with the external states, and every
external contraction contributes the factor $1/2$ of
Eq.~\eqref{eq:half_contraction_rule}.  Hence
\begin{equation}
\boxed{
\begin{aligned}
\mathscr M^{(2),\mathrm{point}}_{
\lambda_3\lambda_4;\lambda_1\lambda_2}(3,4;1,2)
={}&\frac14(2\pi)^3\delta^{(3)}
\!\left(\vec k_1+\vec k_2-\vec k_3-\vec k_4\right)
\\
&\times
\left[\mathcal B_{R,\lambda_3\lambda_4;
\lambda_1\lambda_2}(3,4|1,2)\right]_{(34)(12)}.
\end{aligned}
}
\label{eq:two_to_two_point_matrix_element}
\end{equation}
Thus the four-leg point diagram is not zero merely because the external
momenta are distinct.  Its universal momentum restriction is the spatial
delta function in Eq.~\eqref{eq:two_to_two_point_matrix_element}.

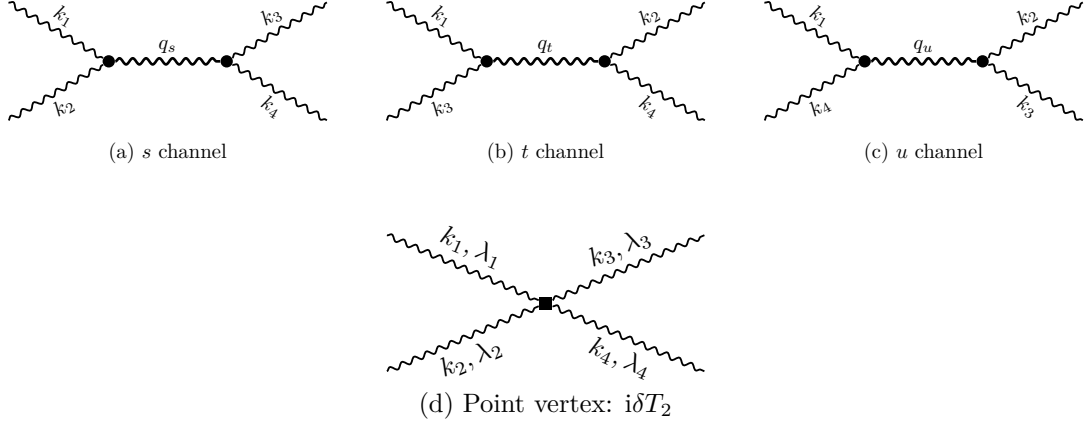
\begin{figure}[t]
\centering
\begin{tikzpicture}[
  ext/.style={decorate,decoration={snake,amplitude=1.0pt,segment length=5pt},
    line width=0.7pt},
  int/.style={decorate,decoration={snake,amplitude=1.15pt,segment length=5pt},
    line width=1.0pt},
  cubic/.style={circle,fill=black,inner sep=2.1pt},
  quartic/.style={rectangle,fill=black,inner sep=2.4pt},
  every node/.style={font=\small}
]

\begin{scope}[shift={(0,0)}]
  \node[quartic] (v) at (0,0) {};
  \draw[ext] (-2.1,0.9) -- node[above,sloped] {$k_1,\lambda_1$} (v);
  \draw[ext] (-2.1,-0.9) -- node[below,sloped] {$k_2,\lambda_2$} (v);
  \draw[ext] (v) -- node[above,sloped] {$k_3,\lambda_3$} (2.1,0.9);
  \draw[ext] (v) -- node[below,sloped] {$k_4,\lambda_4$} (2.1,-0.9);
  \node at (0,-1.35) {(d) Point vertex: $\ii\delta T_2$};
\end{scope}

\begin{scope}[shift={(-5.0,3.2)},scale=0.78,transform shape]
  \node[cubic] (l) at (-1.0,0) {};
  \node[cubic] (r) at (1.0,0) {};
  \draw[ext] (-2.7,1.0) -- node[above,sloped] {$k_1$} (l);
  \draw[ext] (-2.7,-1.0) -- node[below,sloped] {$k_2$} (l);
  \draw[int] (l) -- node[above] {$q_s$} (r);
  \draw[ext] (r) -- node[above,sloped] {$k_3$} (2.7,1.0);
  \draw[ext] (r) -- node[below,sloped] {$k_4$} (2.7,-1.0);
  \node at (0,-1.55) {(a) $s$ channel};
\end{scope}

\begin{scope}[shift={(0,3.2)},scale=0.78,transform shape]
  \node[cubic] (l) at (-1.0,0) {};
  \node[cubic] (r) at (1.0,0) {};
  \draw[ext] (-2.7,1.0) -- node[above,sloped] {$k_1$} (l);
  \draw[ext] (-2.7,-1.0) -- node[below,sloped] {$k_3$} (l);
  \draw[int] (l) -- node[above] {$q_t$} (r);
  \draw[ext] (r) -- node[above,sloped] {$k_2$} (2.7,1.0);
  \draw[ext] (r) -- node[below,sloped] {$k_4$} (2.7,-1.0);
  \node at (0,-1.55) {(b) $t$ channel};
\end{scope}

\begin{scope}[shift={(5.0,3.2)},scale=0.78,transform shape]
  \node[cubic] (l) at (-1.0,0) {};
  \node[cubic] (r) at (1.0,0) {};
  \draw[ext] (-2.7,1.0) -- node[above,sloped] {$k_1$} (l);
  \draw[ext] (-2.7,-1.0) -- node[below,sloped] {$k_4$} (l);
  \draw[int] (l) -- node[above] {$q_u$} (r);
  \draw[ext] (r) -- node[above,sloped] {$k_2$} (2.7,1.0);
  \draw[ext] (r) -- node[below,sloped] {$k_3$} (2.7,-1.0);
  \node at (0,-1.55) {(c) $u$ channel};
\end{scope}
\end{tikzpicture}
\caption{Tree topologies contributing to the connected $2\to2$ matrix
element.  Panels (a)--(c) are the contracted $s$, $t$, and $u$ pairings
contained in $-\tfrac12\delta T_1^2$.  Panel (d) is the four-leg point
vertex reconstructed from $\delta T_2$.  Every internal line is
an on-shell canonical contraction, not an off-shell covariant propagator.}
\label{fig:two_to_two_tree_topologies}
\end{figure}

The term $-\tfrac12\delta T_1^2$ gives the three tree channels obtained by
canonical contractions through on-shell intermediate states.
Their internal spatial momenta may be chosen as
\begin{equation}
\vec q_s=\vec k_1+\vec k_2,
\qquad
\vec q_t=\vec k_1-\vec k_3,
\qquad
\vec q_u=\vec k_1-\vec k_4,
\qquad
\omega_{q_I}=|\vec q_I|.
\label{eq:two_to_two_channel_momenta}
\end{equation}
The point contribution and the three tree contraction channels are displayed in
Fig.~\ref{fig:two_to_two_tree_topologies}.

Let $\operatorname{Cr}_3$ denote the crossing of leg $3$ defined in
Appendix~\ref{app:two_to_two_crossing}.  The three cyclic contraction channels of
Eq.~\eqref{app:eq:M_three_to_one_kernel_result} become
\begin{align}
\mathcal E_u(3,4|1,2)
&=\operatorname{Cr}_3\mathcal D(4;1|2,3),
\nonumber\\
\mathcal E_t(3,4|1,2)
&=\operatorname{Cr}_3\mathcal D(4;2|3,1),
\nonumber\\
\mathcal E_s(3,4|1,2)
&=\operatorname{Cr}_3\mathcal D(4;3|1,2).
\label{eq:two_to_two_crossed_contractions}
\end{align}
Here $\mathcal E_I$ contains the common spatial delta function, the
on-shell factor $1/(2\omega_{q_I})$, and the intermediate-helicity sum.
Crossing changes the relative sign of the contraction contribution, so that
\begin{equation}
\boxed{
\mathscr M^{(2)}_{2\to2}
=
\mathscr M^{(2),\mathrm{point}}_{2\to2}
-\frac14\left(\mathcal E_s+\mathcal E_t+\mathcal E_u\right).
}
\label{eq:two_to_two_complete_tree_kernel}
\end{equation}
This is the lowest-order resolution of the blob in
Fig.~\ref{fig:two_to_two_complete_blob}.

The decay selection rule is transported by crossing.  For generic
non-collinear data, the helicity sectors allowed by the selection rules are
\begin{equation}
\boxed{
++\to++,
\qquad
--\to--,
\qquad
+-\to+-,
\qquad
+-\to-+,
}
\label{eq:two_to_two_helicity_selection_rule}
\end{equation}
together with Bose-equivalent orderings.  Equivalently,
\begin{equation}
\boxed{
\begin{aligned}
\mathscr M^{(2)}_{\lambda_3\lambda_4;++}&=0
&&\text{unless }(\lambda_3,\lambda_4)=(+,+),
\\
\mathscr M^{(2)}_{\lambda_3\lambda_4;--}&=0
&&\text{unless }(\lambda_3,\lambda_4)=(-,-),
\\
\mathscr M^{(2)}_{++;+-}
=\mathscr M^{(2)}_{--;+-}&=0.
\end{aligned}
}
\label{eq:two_to_two_vanishing_helicity_families}
\end{equation}
Here ``allowed'' means that the corresponding sector is not eliminated by
the helicity selection rules; it does not assign a kinematics-independent
numerical value to the matrix element.  Once the external momenta and, when
appropriate, the wave-packet profiles have been specified, the remaining
angular Green-function and principal-value integrals are evaluated
numerically.  Special kinematic configurations may therefore give zero even
within an allowed helicity sector.
The signed-momentum calculation is given in
Appendix~\ref{app:two_to_two_crossing}.

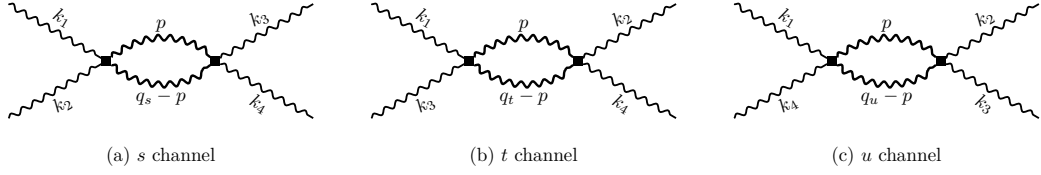
\begin{figure}[t]
\centering
\begin{tikzpicture}[
  ext/.style={decorate,decoration={snake,amplitude=1.0pt,segment length=5pt},
    line width=0.7pt},
  int/.style={decorate,decoration={snake,amplitude=1.1pt,segment length=4.5pt},
    line width=0.9pt},
  quartic/.style={rectangle,fill=black,inner sep=2.4pt},
  every node/.style={font=\small}
]
\begin{scope}[shift={(-4.8,0)},scale=0.72,transform shape]
  \node[quartic] (l) at (-1.0,0) {};
  \node[quartic] (r) at (1.0,0) {};
  \draw[ext] (-2.8,1.05) -- node[above,sloped] {$k_1$} (l);
  \draw[ext] (-2.8,-1.05) -- node[below,sloped] {$k_2$} (l);
  \draw[int] (l) to[bend left=38] node[above] {$p$} (r);
  \draw[int] (l) to[bend right=38] node[below] {$q_s-p$} (r);
  \draw[ext] (r) -- node[above,sloped] {$k_3$} (2.8,1.05);
  \draw[ext] (r) -- node[below,sloped] {$k_4$} (2.8,-1.05);
  \node at (0,-1.75) {(a) $s$ channel};
\end{scope}

\begin{scope}[shift={(0,0)},scale=0.72,transform shape]
  \node[quartic] (l) at (-1.0,0) {};
  \node[quartic] (r) at (1.0,0) {};
  \draw[ext] (-2.8,1.05) -- node[above,sloped] {$k_1$} (l);
  \draw[ext] (-2.8,-1.05) -- node[below,sloped] {$k_3$} (l);
  \draw[int] (l) to[bend left=38] node[above] {$p$} (r);
  \draw[int] (l) to[bend right=38] node[below] {$q_t-p$} (r);
  \draw[ext] (r) -- node[above,sloped] {$k_2$} (2.8,1.05);
  \draw[ext] (r) -- node[below,sloped] {$k_4$} (2.8,-1.05);
  \node at (0,-1.75) {(b) $t$ channel};
\end{scope}

\begin{scope}[shift={(4.8,0)},scale=0.72,transform shape]
  \node[quartic] (l) at (-1.0,0) {};
  \node[quartic] (r) at (1.0,0) {};
  \draw[ext] (-2.8,1.05) -- node[above,sloped] {$k_1$} (l);
  \draw[ext] (-2.8,-1.05) -- node[below,sloped] {$k_4$} (l);
  \draw[int] (l) to[bend left=38] node[above] {$p$} (r);
  \draw[int] (l) to[bend right=38] node[below] {$q_u-p$} (r);
  \draw[ext] (r) -- node[above,sloped] {$k_2$} (2.8,1.05);
  \draw[ext] (r) -- node[below,sloped] {$k_3$} (2.8,-1.05);
  \node at (0,-1.75) {(c) $u$ channel};
\end{scope}
\end{tikzpicture}
\caption{One-loop bubble diagrams generated by
$-\tfrac12\delta T_2^2$.  The two internal lines carry momenta
$\vec p$ and $\vec q_I-\vec p$.  The three panels differ only in the
attachment of the external legs.}
\label{fig:two_to_two_T2_squared_channels}
\end{figure}

The same $s$, $t$, and $u$ classification applies at loop order.  For any
fixed topology, the external attachments are
\begin{equation}
s:(1,2)|(3,4),
\qquad
t:(1,3)|(2,4),
\qquad
u:(1,4)|(2,3).
\label{eq:two_to_two_loop_channel_pairings}
\end{equation}
The first loop contribution containing two four-leg vertices is
\begin{equation}
\boxed{
\left.
\mathscr M_{\lambda_3\lambda_4;\lambda_1\lambda_2}
\right|_{\delta T_2^2}
=
-\frac12
\left\langle k_3,\lambda_3;k_4,\lambda_4\right|
\delta T_2^2
\left|k_1,\lambda_1;k_2,\lambda_2\right\rangle_{\rm conn}.
}
\label{eq:two_to_two_T2_squared_matrix_element}
\end{equation}
For a channel $I=s,t,u$, the two internal lines may be assigned momenta
$\vec p$ and $\vec q_I-\vec p$, as shown in
Fig.~\ref{fig:two_to_two_T2_squared_channels}.

Because the exponential in Eq.~\eqref{eq:S_two_reconstructed_generators}
is not truncated, the two reconstructed vertices generate arbitrarily many
insertions.  If $V_3$ and $V_4$ are the numbers of cubic and four-leg
vertices in a connected $2\to2$ diagram, then
\begin{equation}
I=\frac{3V_3+4V_4-4}{2},
\qquad
\boxed{
L=I-(V_3+V_4)+1
=\frac{V_3+2V_4-2}{2}.
}
\label{eq:two_to_two_loop_count}
\end{equation}
Thus the same formal exponential generates the tree diagrams and the
successive loop topologies constructed from the vertices already known.

%=====================================================================
\section{Discussion and conclusions}
\label{sec:conclusion}
%=====================================================================

The central result of this work is Theorem~3: within the polynomial
bosonic CCR algebra, preservation of the outgoing canonical commutation
relations order by order is sufficient to solve the inverse adjoint
problem constructively.  The CCR hierarchy therefore serves both as a
consistency test for a perturbative outgoing map and as its integrability
condition.  Once the independently
computed coefficients
\[
a_i^{\rm out}
=
a_i+\sum_{n\geq2}\varepsilon^{n-1}\delta a_{n,i}^{\rm out}
\]
satisfy the outgoing CCRs, the coefficients of a self-adjoint BCH
generator can be reconstructed explicitly and recursively, with no
independent ansatz for the scattering operator.

At order \(n\), the generators already reconstructed at lower
orders determine the complete BCH contribution \(I_{n,i}\).  Subtracting
it from the independently obtained outgoing coefficient defines
\begin{equation}
R_{n,i}
=
\delta a_{n,i}^{\rm out}-I_{n,i}.
\label{eq:discussion_residual_reconstruction}
\end{equation}
The content of Theorem~3 is that the outgoing CCR hierarchy forces this
residual to satisfy the homogeneous integrability
conditions of Theorem~1.  No additional integrability condition is
required; the new coefficient follows from
\begin{equation}
\boxed{
\delta T_{n-1}
=-2\ii\sum_d\sum_{r=0}^{d-1}\frac{1}{r+1}
\int_i a^\dagger(i)R_{n,[d]}^{(r,d-r)}(i),
}
\label{eq:discussion_explicit_generator}
\end{equation}
and obeys
\begin{equation}
\boxed{
\ii[a_i,\delta T_{n-1}]
=
R_{n,i}.
}
\label{eq:final_NSF_reconstruction}
\end{equation}
Consequently,
\begin{equation}
\boxed{
\text{outgoing CCRs}
\quad\Longrightarrow\quad
\text{explicit recursive reconstruction of all }\delta T_n
\quad\Longrightarrow\quad
S=e^{\ii\delta T}.
}
\label{eq:discussion_CCR_to_S}
\end{equation}

This implication is particularly useful for the perturbative NSF program
developed in Refs.~\cite{QSI,QSII,QSIII}.  The coefficients
\(\delta a_n^{\rm out}\) are obtained there directly from the gravitational
field equations, null-surface reconstruction, and asymptotic matching,
without first postulating the operator \(S\).  The present construction
provides an algebraically independent check of those
calculations.  If the resulting outgoing operators fail the perturbative
CCR hierarchy, the recursive reconstruction stops.  Conversely, if they
satisfy the CCRs, Theorem~3 guarantees that the corresponding
\(\delta T_n\) exist within the stated algebraic class and determines them
explicitly.  Unlike an a posteriori comparison with a BCH ansatz,
preservation of the CCRs is sufficient to reconstruct
the generator itself.

For the NSF coefficients, the complete ordered outgoing coefficient
\(\delta a_n^{\rm out}\) is tested directly against the CCR hierarchy.
Only afterward is the BCH contribution fixed by lower orders subtracted,
\(R_n=\delta a_n^{\rm out}-I_n\), for the sole purpose of reconstructing
the new generator coefficient.  The residual \(R_n\) is not an additional
correction to the NSF operator.  After the reconstructed coefficients are
assembled into the self-adjoint phase and exponentiated, the formal
asymptotic map is obtained order by order.
Once \(S=\exp(\ii\delta T)\) has been reconstructed,
\(a^{\rm out}=S^\dagger a^{\rm in}S\) preserves the canonical
algebra identically as a formal series through the reconstructed orders,
and no separate perturbative test of
\(S^\dagger S=\mathbf1\) is required.
At the first nonlinear order, the two direct CCR tests for
\(\delta a_2\) are satisfied.  At order three, the
operator-degree-zero sector and all four independent operator-degree-two
sectors are also satisfied.  The scalar mixed identity is now obtained
directly from the NSF cubic kernel.  The three inequivalent placements of
the single creation operator are parameterized by \(c=0,1,2\); normal
ordering produces a linear contraction \(D_c=cD_1\), and the explicit
calculation gives
\begin{equation}
E^{a\bar a;(0)}_{3,ij}(c)
=
\frac{1-c}{2}
\int_{rs}K(i;r,s)K(j;r,s)^*.
\label{eq:discussion_scalar_ordering_result}
\end{equation}
Hence the Ashtekar CCR selects the symmetric ordering \(c=1\), for which
the linear contraction of \(\delta a_3\) cancels identically the
double-contraction source from
\([\delta a_2,\delta a_2^\dagger]\).  This is not a condition imposed on
\(D\).  It is an identity following from the explicit cubic NSF kernel and
the already verified order-two relations.  The no-dagger and mixed
degree-two results are displayed respectively in
Eqs.~\eqref{eq:app_E3_aa_degree2_test} and
\eqref{eq:app_E3_mixed_degree2_test}.  Their evaluation retains the full
momentum distributions, mode normalizations, helicity Kronecker deltas,
principal-value prescriptions, and relative contraction multiplicities.
Thus the canonically ordered flat solution-space NSF coefficient passes the
direct outgoing Ashtekar-CCR test through \(\delta a_3\), i.e. through
order \(\varepsilon^2\).

Here ``flat'' does not mean that NSF begins with an external Minkowski
background.  The variables \(x^a\)
are labels of the four-dimensional solution space of the cut equation, and
the spacetime geometry is reconstructed on that space.  The specialization
proved here consists in taking the metric reconstructed on the solution
space to be \(\eta_{ab}\) and keeping the corresponding cone kinematics
fixed.  In this sector the graviton scattering
map generated by the null-surface equations preserves the Ashtekar CCR
through \(\delta a_3\).  By Theorem~3 this map admits, to that
perturbative order, a formal implementation by a self-adjoint BCH generator
and its unitary exponential.  This is a statement about the flat
solution-space NSF theory itself; comparison with the conventional
Poincar\'e/Weinberg scattering sector is a subsequent kinematical
restriction, not the definition of the construction.

The ordering result also has a direct consequence for the four-graviton
matrix element constructed from two third-order outgoing coefficients.
For the canonical choice \(c=1\),
\(\delta a_3=\delta a_3^{(3)}+\delta a_3^{(1)}\), so the product
\(\delta a_3\delta a_3\) contains the four operator-degree sectors
\((3,3)\), \((3,1)\), \((1,3)\), and \((1,1)\).  Their explicit
contractions are given in
Sec.~\ref{subsec:direct_deltaa3_deltaa3_sectors}.  The cubic--cubic term is
the irreducible one-loop four-leg kernel.  The linear--cubic term generated
by the ordering self-contraction is connected but one-particle reducible,
whereas the linear--linear term factorizes into two one-particle
corrections and is absent from the connected amplitude.  The linear term
required by the Ashtekar CCR therefore has a definite scattering
interpretation without modifying the irreducible cubic--cubic loop.

The generators have a direct vertex interpretation.  The reconstructed
\(\delta T_1\) supplies the three-graviton vertex; once the cubic residual
passes its compatibility test, \(\delta T_2\) supplies the four-graviton
vertex.  The cubic residual fixes the quartic sector
\(\delta T_2^{(4)}\), whereas the scalar CCR controls compatibility and
self-adjointness of the bilinear sector \(\delta T_2^{(2)}\) needed for the
complete generator.  Products of these operators are not new inputs: their
relative coefficients follow from the exponential.  In particular, the
leading \(3\to1\) matrix element contains a four-leg term from
\(\ii\delta T_2\) and three terms obtained by canonical contractions
through on-shell intermediate states from \(-\delta T_1^2/2\).  For
generic non-collinear data, both the contracted terms
and the point-kernel contribution vanish in the incoming equal-helicity
sectors \(+++\) and \(---\).  Hence the complete equal-helicity amplitudes
vanish.  The mixed-helicity sectors are not excluded by this selection
rule; their values depend on the external kinematics through
\(\mathcal A_R\).  The internal momenta of the contracted terms are fixed by the
spatial delta functions, so no loop integration occurs in this process.

Crossing the same four-leg structure gives the connected \(2\to2\)
kernel.  At the lowest nontrivial order it is the sum of the point term
generated by \(\ii\delta T_2\) and the \(s\)-, \(t\)-, and \(u\)-channel
canonical contractions through on-shell intermediate states generated by
\(-\delta T_1^2/2\).  The point contribution
does not vanish for distinct external momenta; its support is the overall
spatial momentum delta function.  Using the established equal-helicity
kernel identities and crossing, the generic helicity sectors left open by
the selection rules are
\(++\to++\), \(--\to--\), \(+-\to+-\), and \(+-\to-+\), together with
Bose-equivalent orderings.  Whether an allowed sector vanishes at a
particular kinematic point is determined only after the remaining angular
Green-function and principal-value integrals have been evaluated.

The exponential of the two reconstructed vertices also generates an
infinite sequence of connected topologies.  For \(2\to2\) scattering the
loop number obeys Eq.~\eqref{eq:two_to_two_loop_count}; for example,
\(-\delta T_2^2/2\) produces the one-loop bubble in the three external
pairings.  These are on-shell canonical contractions rather than diagrams
built from off-shell covariant propagators.  The construction determines
their operator kernels and combinatorial weights, while numerical values require
specified external momenta or wave packets and the evaluation of the
remaining angular and principal-value integrals.  The topological loop
count therefore neither establishes equivalence with conventional
off-shell Feynman loops nor, by itself, proves ultraviolet finiteness.

The explicit third-order residual constructed here uses the flat
solution-space specialization of Ref.~\cite{QSII} with \(U=\mathbf1\):
the solution space labelled by \(x^a\) is endowed with
\(\eta_{ab}\), and the associated cut, geodesic, affine and cone-measure
data are held at their flat values.  There is no recursive-cut
contribution in this sector.  Geodesic and metric corrections must be
included together when extending the calculation beyond that
specialization.

\appendix
\numberwithin{equation}{section}

%=====================================================================
\section{Perturbative verification of the outgoing CCRs}
\label{app:compatibility_conditions}
%=====================================================================

This appendix contains the direct canonical test of the NSF coefficients.
Every kernel below is a complete distribution: its momentum delta function,
mode normalization, and helicity labels are included.  In particular, each
contraction contains the Kronecker delta through
\begin{equation}
[a_{\lambda_i}(\vec k_i),a_{\lambda_j}^\dagger(\vec k_j)]
=\omega_i\,\delta_{\lambda_i\lambda_j}
\delta^{(3)}(\vec k_i-\vec k_j)
\equiv\Delta(i,j).
\label{eq:app_explicit_helicity_CCR}
\end{equation}

%---------------------------------------------------------------------
\subsection{The coefficient \texorpdfstring{\(\delta a_2\)}{delta a2}}
\label{app:quadratic_compatibility}
%---------------------------------------------------------------------

Write the already normally ordered quadratic coefficient as
\begin{equation}
\delta a_{2,i}
=\int_{12}K(i;1,2)a(1)a(2)
+\int_{12}L(i,1|2)a^\dagger(1)a(2),
\qquad K(i;1,2)=K(i;2,1).
\label{eq:app_deltaa2_KL}
\end{equation}
The two direct order-two CCR expressions are
\begin{align}
E^{aa}_{2,ij}
&\equiv[\delta a_{2,i},a_j]+[a_i,\delta a_{2,j}],
\label{eq:app_E2_aa_def}
\\
E^{a\bar a}_{2,ij}
&\equiv[\delta a_{2,i},a_j^\dagger]
+[a_i,\delta a_{2,j}^\dagger].
\label{eq:app_E2_mixed_def}
\end{align}
Using Eq.~\eqref{eq:half_contraction_rule}, the first expression is
\begin{equation}
E^{aa}_{2,ij}
=\frac12\int_r
\bigl[L(j,i|r)-L(i,j|r)\bigr]a(r).
\label{eq:app_E2_aa_explicit}
\end{equation}
The mixed expression separates into one-annihilator and one-creator sectors,
\begin{equation}
\begin{aligned}
E^{a\bar a}_{2,ij}
={}&\int_r
\left[K(i;j,r)+\frac12L(j,r|i)^*\right]a(r)
\\
&+\int_r
\left[\frac12L(i,r|j)+K(j;i,r)^*\right]a^\dagger(r).
\end{aligned}
\label{eq:app_E2_mixed_explicit}
\end{equation}
Consequently, the two order-two CCRs are equivalent to
\begin{equation}
\boxed{
L(i,j|r)=L(j,i|r),
\qquad
K(i;j,r)=-\frac12L(j,r|i)^*.
}
\label{eq:app_E2_kernel_relations}
\end{equation}
The complete NSF kernels obey both relations in
Eq.~\eqref{eq:app_E2_kernel_relations}; hence
\begin{equation}
\boxed{E^{aa}_{2,ij}=0,\qquad E^{a\bar a}_{2,ij}=0.}
\label{eq:app_E2_verified}
\end{equation}
These are the first two direct checks.  They involve only \(\delta a_2\)
and the incoming oscillator algebra.

%---------------------------------------------------------------------
\subsection{The coefficient \texorpdfstring{\(\delta a_3\)}{delta a3}}
\label{app:cubic_compatibility}
%---------------------------------------------------------------------

After normal ordering, separate the cubic and one-operator pieces,
\begin{equation}
\delta a_{3,i}=\delta a_{3,i}^{(3)}+\delta a_{3,i}^{(1)},
\label{eq:app_deltaa3_degree_split}
\end{equation}
with
\begin{align}
\delta a_{3,i}^{(3)}
={}&\int_{123}A(i;1,2,3)a(1)a(2)a(3)
\nonumber\\
&+\int_{123}B(i,1|2,3)a^\dagger(1)a(2)a(3)
\nonumber\\
&+\int_{123}C(i,1,2|3)a^\dagger(1)a^\dagger(2)a(3),
\label{eq:app_deltaa3_ABC}
\\
\delta a_{3,i}^{(1)}
={}&\int_r\left[D(i|r)a(r)+\widetilde D(i|r)a^\dagger(r)\right].
\label{eq:app_deltaa3_D}
\end{align}
Here \(A\), \(B\), and \(C\) are symmetric in their three
annihilator, two annihilator, and two creator slots, respectively.  They
are the kernels inherited from the classically computed \(\delta a_3\)
after the chosen operator ordering.

The direct order-three CCR expressions are
\begin{align}
E^{aa}_{3,ij}
\equiv{}&[\delta a_{3,i},a_j]+[a_i,\delta a_{3,j}]
+[\delta a_{2,i},\delta a_{2,j}],
\label{eq:app_E3_aa_def}
\\
E^{a\bar a}_{3,ij}
\equiv{}&[\delta a_{3,i},a_j^\dagger]
+[a_i,\delta a_{3,j}^\dagger]
+[\delta a_{2,i},\delta a_{2,j}^\dagger].
\label{eq:app_E3_mixed_def}
\end{align}
These equations contain terms of operator degree zero and two.

\subsubsection{Operator degree zero: explicit ordering identity}

The scalar part of the first CCR is
\begin{equation}
E^{aa;(0)}_{3,ij}
=\frac12\left[\widetilde D(j|i)-\widetilde D(i|j)\right].
\label{eq:app_E3_aa_degree0}
\end{equation}
The mixed cubic monomial contains one creator and two annihilators.  Its
quantization therefore has an ordering ambiguity which is invisible at the
classical level but becomes a linear operator after the bosonic CCR is
used.  Write the relevant normally ordered monomial as
\(a^\dagger(1)a(2)a(3)\).  The three inequivalent placements of the creator
are
\begin{align}
{\cal O}_L(1|2,3)
&=a^\dagger(1)a(2)a(3),
\label{eq:app_ordering_L}
\\
{\cal O}_M(1|2,3)
&=\frac12\left[
 a(2)a^\dagger(1)a(3)
+a(3)a^\dagger(1)a(2)\right],
\label{eq:app_ordering_M}
\\
{\cal O}_R(1|2,3)
&=a(2)a(3)a^\dagger(1).
\label{eq:app_ordering_R}
\end{align}
Using Eq.~\eqref{eq:app_explicit_helicity_CCR},
\begin{align}
{\cal O}_L
&=a^\dagger(1)a(2)a(3),
\nonumber\\
{\cal O}_M
&=a^\dagger(1)a(2)a(3)
+\frac12\left[\Delta(2,1)a(3)+\Delta(3,1)a(2)\right],
\label{eq:app_ordering_M_normal}
\\
{\cal O}_R
&=a^\dagger(1)a(2)a(3)
+\Delta(2,1)a(3)+\Delta(3,1)a(2).
\label{eq:app_ordering_R_normal}
\end{align}
It is therefore convenient to parameterize all three choices by
\begin{equation}
c=0,1,2,
\qquad
{\cal O}_c
=
a^\dagger(1)a(2)a(3)
+\frac{c}{2}\left[\Delta(2,1)a(3)+\Delta(3,1)a(2)\right],
\label{eq:app_ordering_c_definition}
\end{equation}
where \(c=0\), \(1\), and \(2\) correspond respectively to
Eqs.~\eqref{eq:app_ordering_L},
\eqref{eq:app_ordering_M}, and
\eqref{eq:app_ordering_R}.  No \(a^\dagger\)-linear term is produced by
these contractions, so
\begin{equation}
\widetilde D(i|r)=0,
\qquad
E^{aa;(0)}_{3,ij}\equiv0.
\label{eq:app_tildeD_zero_ordering}
\end{equation}

We now calculate the annihilation-linear kernel rather than imposing its
CCR condition.  The mixed cubic branch of Eq.~\eqref{eq:app_deltaa3_ABC}
is
\begin{equation}
\int_{123}B(i,1|2,3)\,a^\dagger(1)a(2)a(3),
\qquad
B(i,1|2,3)=B(i,1|3,2).
\label{eq:app_B_branch_for_D}
\end{equation}
Substitution of Eq.~\eqref{eq:app_ordering_c_definition} gives
\begin{align}
\delta a_{3,i}^{(1)}(c)
={}&\frac{c}{2}\int_{123}B(i,1|2,3)
\left[\Delta(2,1)a(3)+\Delta(3,1)a(2)\right].
\label{eq:app_deltaa3_linear_explicit_delta}
\end{align}
Using the contraction rule
\(\int_1\Delta(2,1)F(1)=F(2)/2\) and the Bose symmetry of the
annihilator slots, the two terms are equal.  Hence
\begin{equation}
\boxed{
\delta a_{3,i}^{(1)}(c)
=
\int_j D_c(i|j)a(j),
\qquad
D_c(i|j)
=
\frac{c}{2}\int_x B(i,x|x,j).
}
\label{eq:app_Dc_from_B}
\end{equation}
Thus \(D\) is not an independent kernel.  Explicitly,
\begin{equation}
\int_x\equiv
\sum_{\lambda_x}\int\frac{d^3k_x}{2\omega_x},
\qquad
\int_{xr}\equiv
\sum_{\lambda_x,\lambda_r}
\int\frac{d^3k_x}{2\omega_x}
\frac{d^3k_r}{2\omega_r},
\label{eq:app_D_integrals_explicit}
\end{equation}
and all momentum delta distributions carried by the kernels are retained.

For the NSF coefficient the mixed cubic kernel is
\begin{align}
B(i,1|2,3)
={}&\frac12\int_r\Big[
K(i;2,r)L(r,1|3)
+K(i;3,r)L(r,1|2)
\Big]
\nonumber\\
&-\frac14\int_rL(i,1|r)\bar L(2,3|r)
\nonumber\\
&+\frac14\int_r\Big[
L(i,r|3)\bar L(r,2|1)
+L(i,r|2)\bar L(r,3|1)
\Big],
\label{eq:app_B_for_scalar_identity}
\end{align}
where \(\bar L\equiv L^*\).  For the symmetric ordering \(c=1\),
Eqs.~\eqref{eq:app_Dc_from_B} and
\eqref{eq:app_B_for_scalar_identity} give the fully expanded
two-mode integral
\begin{align}
D_1(i|j)
={}&
\frac14\int_{xr}K(i;x,r)L(r,x|j)
+\frac14\int_{xr}K(i;j,r)L(r,x|x)
\nonumber\\
&-\frac18\int_{xr}L(i,x|r)\bar L(x,j|r)
\nonumber\\
&+\frac18\int_{xr}L(i,r|j)\bar L(r,x|x)
+\frac18\int_{xr}L(i,r|x)\bar L(r,j|x).
\label{eq:app_D1_fully_expanded}
\end{align}
At this point only the already verified order-two identities are used:
\begin{equation}
L(x,r|j)=L(r,x|j),
\qquad
K(i;x,r)=-\frac12\bar L(x,r|i).
\label{eq:app_order2_identities_for_D}
\end{equation}
The first two terms of Eq.~\eqref{eq:app_D1_fully_expanded} become
\begin{align}
-\frac18\int_{xr}\bar L(x,r|i)L(x,r|j),
\qquad
-\frac18\int_{xr}\bar L(j,r|i)L(r,x|x).
\label{eq:app_D_first_two_reduced}
\end{align}
In the last term of Eq.~\eqref{eq:app_D1_fully_expanded}, interchange the
dummy labels \(x\leftrightarrow r\) and use the symmetry of \(L\).  It is
then
\begin{equation}
+\frac18\int_{xr}L(i,x|r)\bar L(x,j|r),
\label{eq:app_D_last_relabelled}
\end{equation}
which cancels the third term of
Eq.~\eqref{eq:app_D1_fully_expanded} identically.  For notational
convenience, define
\begin{equation}
X(i|j)
\equiv
\frac18\int_{xr}L(i,r|j)\bar L(r,x|x),
\label{eq:app_X_for_D}
\end{equation}
the remaining result is
\begin{equation}
\boxed{
D_1(i|j)
=
-\frac18\int_{xr}\bar L(x,r|i)L(x,r|j)
+X(i|j)-X(j|i)^*.
}
\label{eq:app_D1_identity_decomposition}
\end{equation}
The last two terms are anti-Hermitian under \(i\leftrightarrow j\).
Consequently,
\begin{equation}
\boxed{
D_1(i|j)+D_1(j|i)^*
=
-\frac14\int_{xr}\bar L(x,r|i)L(x,r|j).
}
\label{eq:app_D1_D1star_explicit}
\end{equation}
But the already established quadratic relation in
Eq.~\eqref{eq:app_E2_kernel_relations} gives
\begin{equation}
\int_{xr}K(i;x,r)K(j;x,r)^*
=
\frac14\int_{xr}\bar L(x,r|i)L(x,r|j),
\label{eq:app_KK_as_LLbar_scalar}
\end{equation}
and therefore
\begin{equation}
\boxed{
D_1(i|j)+D_1(j|i)^*
=
-\int_{xr}K(i;x,r)K(j;x,r)^*.
}
\label{eq:app_D_KK_relation}
\end{equation}
This relation has thus been derived from the explicit NSF cubic kernel and
the already verified order-two identities; it has not been imposed as the
order-three CCR.

For completeness, the source on the right-hand side can itself be seen
directly as a double contraction.  The \(aa\) part of
\(\delta a_{2,i}\) gives
\begin{align}
&\left[
\int_{12}K(i;1,2)a(1)a(2),
\int_{34}K(j;3,4)^*a^\dagger(3)a^\dagger(4)
\right]_{\mathbf 1}
\nonumber\\
&\qquad=
\int_{1234}K(i;1,2)K(j;3,4)^*
\left[\Delta(1,3)\Delta(2,4)
+\Delta(1,4)\Delta(2,3)\right]
\nonumber\\
&\qquad=
\frac12\int_{rs}K(i;r,s)K(j;r,s)^*,
\label{eq:app_deltaa2_double_contraction_explicit}
\end{align}
where the final factor follows from the two equal Wick pairings and the
half-contraction rule.  Likewise,
\begin{equation}
\left[
\int_rD_c(i|r)a(r),a^\dagger(j)
\right]
+
\left[
a(i),\int_rD_c(j|r)^*a^\dagger(r)
\right]
=
\frac12\left[D_c(i|j)+D_c(j|i)^*\right].
\label{eq:app_D_external_contraction_explicit}
\end{equation}
Since \(D_c=cD_1\), the scalar mixed CCR is therefore
\begin{align}
E^{a\bar a;(0)}_{3,ij}(c)
&=
\frac12\left[D_c(i|j)+D_c(j|i)^*\right]
+\frac12\int_{rs}K(i;r,s)K(j;r,s)^*
\nonumber\\
&=
\boxed{
\frac{1-c}{2}
\int_{rs}K(i;r,s)K(j;r,s)^* }.
\label{eq:app_E3_mixed_degree0}
\end{align}
Thus
\begin{equation}
\boxed{
c=1
\quad\Longrightarrow\quad
E^{a\bar a;(0)}_{3,ij}\equiv0,
}
\label{eq:app_E3_scalar_c1_identity}
\end{equation}
whereas \(c=0\) leaves the nonzero double-contraction source and \(c=2\)
reverses it rather than canceling it.  The symmetric ordering is therefore
selected by preservation of the Ashtekar radiative CCR.  Combining this
identity with Eq.~\eqref{eq:app_tildeD_zero_ordering} gives
\begin{equation}
\boxed{
E^{aa;(0)}_{3,ij}=0,
\qquad
E^{a\bar a;(0)}_{3,ij}=0
\quad (c=1).
}
\label{eq:app_E3_degree0_verified}
\end{equation}
This places the scalar relation on exactly the same footing as the
normally ordered degree-two identities below: it is an explicit identity
of the NSF kernels, not a condition imposed on \(\delta a_3\).

\subsubsection{Operator degree two and its suborders}

Define normalized symmetrization by
\begin{equation}
\operatorname{Sym}_{12}F(1,2)
\equiv\frac12\bigl[F(1,2)+F(2,1)\bigr].
\label{eq:app_sym12_definition}
\end{equation}
The degree-two part of the no-dagger CCR separates as
\begin{equation}
E^{aa;(2)}_{3,ij}
=\int_{12}E^{aa;(0,2)}_{3,ij}(1,2)a(1)a(2)
+\int_{12}E^{aa;(1,1)}_{3,ij}(1|2)a^\dagger(1)a(2),
\label{eq:app_E3_aa_degree2_split}
\end{equation}
where
\begin{align}
E^{aa;(0,2)}_{3,ij}(1,2)
={}&\frac12\bigl[B(j,i|1,2)-B(i,j|1,2)\bigr]
\nonumber\\
&+\operatorname{Sym}_{12}\int_r
\bigl[
K(i;1,r)L(j,r|2)-K(j;1,r)L(i,r|2)
\bigr],
\label{eq:app_E3_aa_02}
\\
E^{aa;(1,1)}_{3,ij}(1|2)
={}&C(j,i,1|2)-C(i,j,1|2)
\nonumber\\
&+\frac12\int_r
\bigl[
L(i,1|r)L(j,r|2)-L(j,1|r)L(i,r|2)
\bigr].
\label{eq:app_E3_aa_11}
\end{align}
To display this cancellation explicitly, define the quadratic contraction
appearing in Eq.~\eqref{eq:app_E3_aa_02} by
\begin{equation}
\begin{aligned}
S_{ij}(1,2)
\equiv{}&\operatorname{Sym}_{12}\int_r
\bigl[
K(i;1,r)L(j,r|2)-K(j;1,r)L(i,r|2)
\bigr].
\end{aligned}
\label{eq:app_E3_aa_02_contraction}
\end{equation}
Substitute the complete mixed cubic kernel of
Eq.~\eqref{eq:app_B_composed_KL} and form its antisymmetric combination,
\[
B(j,i|1,2)-B(i,j|1,2).
\]
After symmetrization under \(1\leftrightarrow2\), the first line of
Eq.~\eqref{eq:app_B_composed_KL} gives
\begin{equation}
\left[B(j,i|1,2)-B(i,j|1,2)\right]_{\mathrm{first}}
=-S_{ij}(1,2).
\label{eq:app_B_antisymmetric_first_part}
\end{equation}
The second line is symmetric under \(i\leftrightarrow j\), because
\(L(j,i|r)=L(i,j|r)\), and therefore gives zero.  In the last line use
the order-two relation
\begin{equation}
\bar L(r,1|i)=\bar L(1,r|i)=-2K(i;1,r),
\label{eq:app_Lbar_to_K_for_B_antisymmetry}
\end{equation}
and the analogous relations obtained by exchanging the labels.  Its
antisymmetric part is then
\begin{equation}
\left[B(j,i|1,2)-B(i,j|1,2)\right]_{\mathrm{last}}
=-S_{ij}(1,2).
\label{eq:app_B_antisymmetric_third_part}
\end{equation}
Consequently,
\begin{equation}
\boxed{
B(j,i|1,2)-B(i,j|1,2)=-2S_{ij}(1,2).
}
\label{eq:app_B_antisymmetric_identity}
\end{equation}
Substitution in Eq.~\eqref{eq:app_E3_aa_02} now gives the cancellation
term by term,
\begin{equation}
E^{aa;(0,2)}_{3,ij}(1,2)
=\frac12\bigl[-2S_{ij}(1,2)\bigr]+S_{ij}(1,2)=0.
\label{eq:app_E3_aa_02_explicit_cancellation}
\end{equation}
Hence
\begin{equation}
\boxed{E^{aa;(0,2)}_{3,ij}=0.}
\label{eq:app_E3_aa_02_verified}
\end{equation}

We now verify the creator--annihilator sector.  The already established
order-two relations imply
\begin{equation}
L(i,1|r)=L(1,i|r),
\qquad
K(r;i,1)^*=-\frac12L(i,1|r).
\label{eq:app_KL_relations_for_C}
\end{equation}
All labels in this equation include both momentum and helicity.  In
particular, each internal contraction contains the sum over the intermediate
helicity and the Kronecker delta supplied by the CCR.

The selected two-creator branch of the cone coefficient is
\begin{equation}
C(q,i,1|2)=C_{\rm it}(q,i,1|2).
\label{eq:app_C_iterative_identification}
\end{equation}
It is obtained from the two possible quadratic substitutions, after
normalized Bose symmetrization of the two creator slots:
\begin{align}
C_{\rm it}(q,i,1|2)
={}&\frac14\int_r\Big[
L(q,i|r)L(r,1|2)+L(q,1|r)L(r,i|2)
\Big]
\nonumber\\
&-\frac14\int_r L(q,r|2)L(i,1|r).
\label{eq:app_C_iterated_L}
\end{align}
The first line is produced by the two placements in which the
annihilation leg of the mixed quadratic branch is replaced by its
second-order coefficient.  The last term comes from replacing its creation
leg; Eq.~\eqref{eq:app_KL_relations_for_C} has been used to replace the
two-creator quadratic kernel \(K^*\) by \(L\).  This is precisely the
\((--)\) branch of the cone source
\(\partial\Lambda_1\partial\bar\Lambda_2+
\partial\Lambda_2\partial\bar\Lambda_1\), together with the corresponding
three partitions of \(\delta\Omega_3\).  The two creator labels are
symmetrized as complete bosonic labels, including their helicities.

The momentum distributions in the first orientation have the form
\begin{equation}
L(i,1|r)\propto
\delta^{(3)}(\vec k_i+\vec k_1-\vec k_r),
\qquad
L(j,r|2)\propto
\delta^{(3)}(\vec k_j+\vec k_r-\vec k_2).
\label{eq:app_LL_momentum_support}
\end{equation}
Thus the internal integration leaves the common support
\begin{equation}
\delta^{(3)}
(\vec k_i+\vec k_j+\vec k_1-\vec k_2),
\label{eq:app_C_LL_common_support}
\end{equation}
which is exactly the support of
\(C(j,i,1|2)a_i^\dagger a_1^\dagger a_2\).

Using \(L(x,y|r)=L(y,x|r)\), the first term of
Eq.~\eqref{eq:app_C_iterated_L} cancels when the external labels are
exchanged.  The remaining terms give
\begin{align}
&C(j,i,1|2)-C(i,j,1|2)
\nonumber\\
={}&\frac14\int_r\Big[
L(j,1|r)L(r,i|2)-L(i,1|r)L(r,j|2)
\nonumber\\
&\hspace{28mm}
-L(j,r|2)L(i,1|r)+L(i,r|2)L(j,1|r)
\Big]
\nonumber\\
={}&-\frac12\int_r\Big[
L(i,1|r)L(j,r|2)-L(j,1|r)L(i,r|2)
\Big].
\label{eq:app_C_antisymmetric_identity}
\end{align}
Substitution in Eq.~\eqref{eq:app_E3_aa_11} therefore yields
\begin{equation}
\boxed{E^{aa;(1,1)}_{3,ij}=0.}
\label{eq:app_E3_aa_11_verified}
\end{equation}
Consequently, both degree-two sectors of the no-dagger CCR have been
verified:
\begin{equation}
\boxed{E^{aa;(0,2)}_{3,ij}=0,
\qquad E^{aa;(1,1)}_{3,ij}=0.}
\label{eq:app_E3_aa_degree2_test}
\end{equation}

The mixed CCR separates into three normal-ordered suborders,
\begin{align}
E^{a\bar a;(2)}_{3,ij}
={}&\int_{12}E^{a\bar a;(0,2)}_{3,ij}(1,2)a(1)a(2)
\nonumber\\
&+\int_{12}E^{a\bar a;(1,1)}_{3,ij}(1|2)a^\dagger(1)a(2)
\nonumber\\
&+\int_{12}E^{a\bar a;(2,0)}_{3,ij}(1,2)
a^\dagger(1)a^\dagger(2),
\label{eq:app_E3_mixed_degree2_split}
\end{align}
with two independent coefficients,
\begin{align}
E^{a\bar a;(0,2)}_{3,ij}(1,2)
={}&\frac32A(i;j,1,2)
+\frac12C(j,1,2|i)^*
\nonumber\\
&+\operatorname{Sym}_{12}\int_r
K(i;1,r)L(j,2|r)^*,
\label{eq:app_E3_mixed_02}
\\
E^{a\bar a;(1,1)}_{3,ij}(1|2)
={}&B(i,1|j,2)+B(j,2|i,1)^*
\nonumber\\
&+2\int_r K(i;2,r)K(j;1,r)^*
\nonumber\\
&+\frac12\int_r
\bigl[
L(i,1|r)L(j,2|r)^*
-L(i,r|2)L(j,r|1)^*
\bigr].
\label{eq:app_E3_mixed_11}
\end{align}

To evaluate these expressions, set
\begin{equation}
\bar L(i,1|2)\equiv L(i,1|2)^*.
\label{eq:app_Lbar_definition}
\end{equation}
The selected third-order cone source contains one complete first-order
field and one complete second-order field in each term:
\begin{equation}
\mathcal Q_3
=2\bar\eth\eth\,\delta\Omega_3
+\eta^{ab}\left(
\partial_a\Lambda_1\partial_b\bar\Lambda_2
+\partial_a\Lambda_2\partial_b\bar\Lambda_1
\right).
\label{eq:app_Q3_selected_source}
\end{equation}
Consequently, every cubic oscillator coefficient is obtained by inserting
one full quadratic block in one of the two entries of the quadratic cone
source.  The three partitions of \(\delta\Omega_3\) perform the same
replacement in the conformal-factor part.  After normal ordering and
normalized Bose symmetrization, the required kernels are
\begin{align}
A(q;1,2,3)
={}&-\frac16\int_r\Big[
K(q;1,r)\bar L(2,3|r)
+K(q;2,r)\bar L(1,3|r)
\nonumber\\[-1mm]
&\hspace{38mm}
+K(q;3,r)\bar L(1,2|r)
\Big],
\label{eq:app_A_composed_KL}
\\
C(q,1,2|3)
={}&\frac14\int_r\Big[
L(q,1|r)L(r,2|3)
+L(q,2|r)L(r,1|3)
\nonumber\\[-1mm]
&\hspace{38mm}
-L(q,r|3)L(1,2|r)
\Big],
\label{eq:app_C_composed_LL}
\\
B(q,1|2,3)
={}&\frac12\int_r\Big[
K(q;2,r)L(r,1|3)
+K(q;3,r)L(r,1|2)
\Big]
\nonumber\\
&-\frac14\int_rL(q,1|r)\bar L(2,3|r)
\nonumber\\
&+\frac14\int_r\Big[
L(q,r|3)\bar L(r,2|1)
+L(q,r|2)\bar L(r,3|1)
\Big].
\label{eq:app_B_composed_KL}
\end{align}
These equations apply to the complete distributional kernels.  In
particular, the momentum delta functions, the intermediate helicity sum
and its Kronecker delta, and the common radiative normalization are
understood.  The same composition holds separately for
\(\Omega_{\rm ul}\), \(\Omega_{\rm 2int}\),
\(\Lambda_{\rm LN}\), and \(\Lambda_{\rm NN}\): the outer signed factors
are respectively the quadratic \(F_\Omega\) and \(F_\Lambda\), with one
linear momentum replaced by the second-order compound momentum.

\paragraph{The mixed \((0,2)\) identity.}
Using \(K(i;1,r)=-\bar L(1,r|i)/2\) in
Eq.~\eqref{eq:app_A_composed_KL} gives
\begin{align}
\frac32A(i;j,1,2)
=\frac18\int_r\Big[
&\bar L(j,r|i)\bar L(1,2|r)
+\bar L(1,r|i)\bar L(j,2|r)
\nonumber\\
&+\bar L(2,r|i)\bar L(j,1|r)
\Big].
\label{eq:app_threehalf_A_explicit}
\end{align}
Likewise, Eq.~\eqref{eq:app_C_composed_LL} gives
\begin{align}
\frac12C(j,1,2|i)^*
=\frac18\int_r\Big[
&\bar L(j,1|r)\bar L(r,2|i)
+\bar L(j,2|r)\bar L(r,1|i)
\nonumber\\
&-\bar L(j,r|i)\bar L(1,2|r)
\Big].
\label{eq:app_half_Cstar_explicit}
\end{align}
The first term in Eq.~\eqref{eq:app_threehalf_A_explicit} cancels the
last term in Eq.~\eqref{eq:app_half_Cstar_explicit}.  The remaining
terms form two equal pairs, so that
\begin{align}
\frac32A(i;j,1,2)+\frac12C(j,1,2|i)^*
=\frac14\int_r\Big[
&\bar L(1,r|i)\bar L(j,2|r)
\nonumber\\
&+\bar L(2,r|i)\bar L(j,1|r)
\Big].
\label{eq:app_A_Cstar_sum}
\end{align}
On the other hand,
\begin{align}
\operatorname{Sym}_{12}\int_rK(i;1,r)L(j,2|r)^*
=-\frac14\int_r\Big[
&\bar L(1,r|i)\bar L(j,2|r)
\nonumber\\
&+\bar L(2,r|i)\bar L(j,1|r)
\Big].
\label{eq:app_KL_mixed_02_cancellation}
\end{align}
Equations~\eqref{eq:app_A_Cstar_sum} and
\eqref{eq:app_KL_mixed_02_cancellation} cancel term by term.  Their common
momentum support is
\begin{equation}
\delta^{(3)}
(\vec k_i-\vec k_j-\vec k_1-\vec k_2),
\label{eq:app_mixed_02_common_support}
\end{equation}
and their intermediate helicity labels are identified by the same
Kronecker delta.  Therefore
\begin{equation}
\boxed{E^{a\bar a;(0,2)}_{3,ij}=0.}
\label{eq:app_E3_mixed_02_verified}
\end{equation}

\paragraph{The mixed \((1,1)\) identity.}
Substituting Eq.~\eqref{eq:app_B_composed_KL} in the two \(B\) terms,
using \(L(x,y|r)=L(y,x|r)\), and grouping equal terms gives
\begin{align}
&B(i,1|j,2)+B(j,2|i,1)^*
\nonumber\\
={}&-\frac12\int_r\Big[
\bar L(2,r|i)L(1,r|j)
+L(i,1|r)\bar L(j,2|r)
\nonumber\\[-1mm]
&\hspace{37mm}
-L(i,r|2)\bar L(j,r|1)
\Big].
\label{eq:app_B_Bstar_sum}
\end{align}
The first term on the right-hand side is canceled by
\begin{equation}
2\int_rK(i;2,r)K(j;1,r)^*
=\frac12\int_r\bar L(2,r|i)L(1,r|j),
\label{eq:app_KK_mixed_11_cancellation}
\end{equation}
whereas the remaining two terms are canceled by
\begin{equation}
\frac12\int_r\Big[
L(i,1|r)\bar L(j,2|r)
-L(i,r|2)\bar L(j,r|1)
\Big].
\label{eq:app_LLbar_mixed_11_cancellation}
\end{equation}
All three contributions carry
\begin{equation}
\delta^{(3)}
(\vec k_i+\vec k_1-\vec k_j-\vec k_2)
\label{eq:app_mixed_11_common_support}
\end{equation}
and the same intermediate-helicity Kronecker delta.  It follows that
\begin{equation}
\boxed{E^{a\bar a;(1,1)}_{3,ij}=0.}
\label{eq:app_E3_mixed_11_verified}
\end{equation}

The two-creator coefficient is not independent:
\begin{equation}
E^{a\bar a;(2,0)}_{3,ij}(1,2)
=\left[E^{a\bar a;(0,2)}_{3,ji}(1,2)\right]^*.
\label{eq:app_E3_mixed_20_adjoint}
\end{equation}
Hence both independent mixed quadratic identities are verified:
\begin{equation}
\boxed{E^{a\bar a;(0,2)}_{3,ij}=0,
\qquad E^{a\bar a;(1,1)}_{3,ij}=0.}
\label{eq:app_E3_mixed_degree2_test}
\end{equation}
with the \((2,0)\) equation following by adjunction and exchange of the
external labels.

\paragraph{Status of the four independent relations.}
Collecting the independent degree-two sectors, all four relations have now
been verified:
\begin{equation}
\boxed{
\begin{aligned}
E^{aa;(0,2)}_{3,ij}(1,2)&=0,
&\qquad
E^{aa;(1,1)}_{3,ij}(1|2)&=0,
\\
E^{a\bar a;(0,2)}_{3,ij}(1,2)&=0,
&
E^{a\bar a;(1,1)}_{3,ij}(1|2)&=0.
\end{aligned}
}
\label{eq:app_E3_four_verified_relations}
\end{equation}
The \((2,0)\) mixed relation follows from the third equation by adjunction
and exchange of \(i\) and \(j\), so it is not an additional independent
condition.  Thus the direct order-three outgoing CCR test is complete.

%=====================================================================
\section{Tree-level \texorpdfstring{$3\to1$}{3 to 1} matrix element at
second perturbative order}
\label{app:three_to_one_selection}
%=====================================================================

This appendix records the operator reduction and the helicity selection
rules for the leading transition from three incoming gravitons to one
outgoing graviton.  The quadratic NSF kernels are those obtained in
Ref.~\cite{QSI}, while the fixed-Minkowski-cone cubic sector is taken from
Ref.~\cite{QSII}.  The calculation follows the same pattern as the
$2\to1$ analysis of Ref.~\cite{QSI}: the NSF matching supplies a
three-dimensional momentum delta, the frequency of the outgoing leg is
fixed separately by its positive-energy on-shell condition, and the
helicity selection rule can be read from the operator content before the
angular kernels are evaluated.  The new feature at second perturbative
order is that the quartic kernel of Ref.~\cite{QSII} must be
combined with the three contracted tree channels generated by two quadratic
kernels of Ref.~\cite{QSI}.  All statements below refer to the generic
connected kernel. 

%---------------------------------------------------------------------
\subsection{Operator decomposition of the matrix element}
\label{app:subsec:three_to_one_matrix_element}
%---------------------------------------------------------------------

Let the outgoing leg be $4=(\lambda_4,\vec k_4)$ and the three incoming
legs be $i=(\lambda_i,\vec k_i)$, $i=1,2,3$.  The first nonvanishing
coefficient in the expansion of the scattering operator is
\begin{equation}
 S^{(2)}
 =\ii\delta T_2-\frac12\delta T_1^2.
 \label{app:eq:S_second_order_three_to_one}
\end{equation}
Indeed, the cubic operator $\delta T_1$ cannot saturate four external
legs.  The leading connected matrix element is therefore
\begin{equation}
\boxed{
\begin{aligned}
\mathscr M^{(2)}_{\lambda_4;\lambda_1\lambda_2\lambda_3}
(4;1,2,3)
={}&
\langle0|a(4)
\left(
\ii\delta T_2-\frac12\delta T_1^2
\right)
\\
&\hspace{17mm}\times
a^\dagger(1)a^\dagger(2)a^\dagger(3)|0\rangle_{\rm conn}.
\end{aligned}
}
\label{app:eq:M_three_to_one_definition}
\end{equation}

For this matrix element it is convenient to use the two-annihilator
representation of the first generator.  Define
\begin{equation}
 \mathcal T_1
 \equiv
 -2\ii
 \int_4\int_1\int_2
 \mathcal K^{aa}(4;1,2)
 a^\dagger(4)a(1)a(2),
 \qquad
 \delta T_1=\mathcal T_1+\mathcal T_1^\dagger.
 \label{app:eq:Tcal1_aa_definition}
\end{equation}
Only $\mathcal T_1^2$ can have the net operator number required by a
$3\to1$ transition.  In the quartic generator, only the
Hermitian conjugate of the first line of
Eq.~\eqref{eq:deltaT2_general_kernel_form} contributes.  Its contribution
to $\ii\delta T_2$ was written in the previous reconstruction as
\begin{equation}
\left.\ii\delta T_2\right|_{a^\dagger aaa}
=
2\int_4\int_1\int_2\int_3
\mathcal A_R(4;1,2,3)
a^\dagger(4)a(1)a(2)a(3).
\label{app:eq:iT2_relevant_three_to_one}
\end{equation}
The $a^{\dagger 2}a^2$ sector proportional to $\mathcal B_R$ has vanishing
matrix element between a three-particle state and a one-particle state.

To normal order $\mathcal T_1^2$, one annihilator in the left factor must be
contracted with the creator in the right factor.  The two possible
contractions are equal because
$\mathcal K^{aa}(4;1,2)=\mathcal K^{aa}(4;2,1)$, and each contraction
uses the rule~\eqref{eq:half_contraction_rule}.  Define
\begin{equation}
 \mathcal D(4;i|j,k)
 \equiv
 \int_r
 \mathcal K^{aa}(4;i,r)\mathcal K^{aa}(r;j,k).
 \label{app:eq:D_three_to_one_definition}
\end{equation}
After symmetrizing the three annihilator labels, the contracted term is
\begin{equation}
\left.-\frac12\delta T_1^2\right|_{a^\dagger aaa}
=
\frac{2}{3}
\int_4\int_1\int_2\int_3
\sum_{\mathrm{cyc}(1,2,3)}
\mathcal D(4;1|2,3)
a^\dagger(4)a(1)a(2)a(3).
\label{app:eq:T1_squared_relevant_three_to_one}
\end{equation}
Here and below
\begin{equation}
 \sum_{\mathrm{cyc}(1,2,3)}(123)
 \equiv
 (123)+(231)+(312).
 \label{app:eq:cyclic_three_labels}
\end{equation}
Combining Eqs.~\eqref{app:eq:iT2_relevant_three_to_one} and
\eqref{app:eq:T1_squared_relevant_three_to_one}, the relevant operator
coefficient is
\begin{equation}
\boxed{
\left.S^{(2)}\right|_{a^\dagger aaa}
=
2\int_4\int_1\int_2\int_3
\left[
\mathcal A_R(4;1,2,3)
+\frac13
\sum_{\mathrm{cyc}(1,2,3)}
\mathcal D(4;1|2,3)
\right]
a^\dagger(4)a(1)a(2)a(3).
}
\label{app:eq:S2_relevant_three_to_one}
\end{equation}
The four external contractions and the Bose sum over the three incoming
legs then give
\begin{equation}
\boxed{
\mathscr M^{(2)}_{3\to1}(4;1,2,3)
=
\frac34\mathcal A_R(4;1,2,3)
+\frac14
\sum_{\mathrm{cyc}(1,2,3)}
\mathcal D(4;1|2,3).
}
\label{app:eq:M_three_to_one_kernel_result}
\end{equation}
All helicity labels are contained in the complete distributional kernels on
the right-hand side.

%---------------------------------------------------------------------
\subsection{Momentum support and tree topology}
\label{app:subsec:three_to_one_no_loop}
%---------------------------------------------------------------------

Factor the spatial delta functions according to
\begin{align}
 \mathcal K^{aa}(p;i,j)
 &=
 \delta^{(3)}(\vec p-\vec k_i-\vec k_j)
 \kappa^{aa}(p;i,j),
 \label{app:eq:Kaa_reduced_kappa}
 \\
 \mathcal A_R(4;1,2,3)
 &=
 \delta^{(3)}
 (\vec k_4-\vec k_1-\vec k_2-\vec k_3)
 \alpha(4;1,2,3).
 \label{app:eq:A_reduced_alpha}
\end{align}
For the $123$ channel, the two spatial deltas in
Eq.~\eqref{app:eq:D_three_to_one_definition} fix
\begin{equation}
 \vec r_{23}=\vec k_2+\vec k_3,
 \qquad
 \omega_{23}=|\vec r_{23}|,
 \qquad
 r_{23}=(\omega_{23},\vec r_{23}).
 \label{app:eq:r23_on_shell}
\end{equation}
Consequently,
\begin{equation}
\begin{aligned}
 \mathcal D(4;1|2,3)
 ={}&
 \delta^{(3)}
 (\vec k_4-\vec k_1-\vec k_2-\vec k_3)
 \\
 &\times
 \sum_{\lambda_r}
 \frac{
 \kappa^{aa}(4;1,r_{23})
 \kappa^{aa}(r_{23};2,3)
 }{2\omega_{23}}.
\end{aligned}
\label{app:eq:D123_reduced}
\end{equation}
The other two channels follow by the cyclic replacements
$123\to231\to312$.  There is no unfixed momentum integration.  In graph
language the contraction contribution has two vertices and one internal line,
so that
\begin{equation}
 L=I-V+1=1-2+1=0.
 \label{app:eq:three_to_one_loop_count}
\end{equation}
It is a tree contribution, not a loop correction.  The corresponding
$\delta T_2$ and contracted topologies are displayed in
Fig.~\ref{fig:three_to_one_T2_T1_squared}; the remainder of this
appendix reduces their momentum support and helicity content explicitly.

The external support exhibits precisely the kinematical pattern found for
the $2\to1$ transition in Ref.~\cite{QSI}:
\begin{equation}
 \vec k_4=\vec k_1+\vec k_2+\vec k_3,
 \qquad
 \omega_4=|\vec k_4|
 =|\vec k_1+\vec k_2+\vec k_3|.
 \label{app:eq:three_to_one_kinematic_support}
\end{equation}
There is no independent delta function imposing
$\omega_4=\omega_1+\omega_2+\omega_3$.  Instead, the triangle inequality
gives
\begin{equation}
 \boxed{
 \omega_4
 \leq
 \omega_1+\omega_2+\omega_3,
 }
 \label{app:eq:three_to_one_energy_inequality}
\end{equation}
with equality only when the three nonzero incoming spatial momenta are
parallel and point in the same direction.  Thus the incoming spatial
momenta are otherwise unrestricted by an energy-conservation condition,
and for generic non-collinear data one has the strict Bondi-energy balance
\begin{equation}
 \Delta E_{\rm Bondi}
 \equiv
 \omega_4-(\omega_1+\omega_2+\omega_3)<0.
 \label{app:eq:three_to_one_Bondi_energy}
\end{equation}
This is the direct $3\to1$ analogue of the Bondi-energy interpretation of
the $2\to1$ process in Ref.~\cite{QSI}; four-dimensional energy
conservation is recovered only on the exceptional common-collinear
support.

%---------------------------------------------------------------------
\subsection{Helicity constraints on the contracted contribution}
\label{app:subsec:three_to_one_contracted_helicity}
%---------------------------------------------------------------------

The generic bilocal two-annihilator kernel found at quadratic order has the
operator content
\begin{equation}
 \left.\delta a_2\right|_{aa,\mathrm{biloc}}
 \sim
 \kappa^{aa}_{+-}a_+a_-
 +
 \kappa^{aa}_{-+}a_-a_+,
 \label{app:eq:QSI_mixed_aa_rule}
\end{equation}
and therefore
\begin{equation}
\boxed{
 \kappa^{aa}_{\lambda;++}=0,
 \qquad
 \kappa^{aa}_{\lambda;--}=0,
 \qquad \lambda=\pm.
}
\label{app:eq:QSI_equal_helicity_aa_zero}
\end{equation}
Applying this rule at both quadratic vertices gives
\begin{equation}
\begin{array}{c|c|c}
\text{channel}
&\text{condition on external helicities}
&\text{internal helicity}
\\ \hline
123&\lambda_2=-\lambda_3&\lambda_{r_{23}}=-\lambda_1
\\
231&\lambda_3=-\lambda_1&\lambda_{r_{31}}=-\lambda_2
\\
312&\lambda_1=-\lambda_2&\lambda_{r_{12}}=-\lambda_3
\end{array}
\label{app:eq:three_to_one_channel_helicity_table}
\end{equation}
Thus all three contraction channels vanish for $+++$ and for $---$.  For a
mixed incoming triple exactly two cyclic channels are nonzero, and the internal
helicity is fixed to the helicity that occurs only once among the three
incoming legs.  In particular, the discrete internal-helicity sum in
Eq.~\eqref{app:eq:D123_reduced} also collapses.

This is again the $2\to1$ selection pattern of Ref.~\cite{QSI}, now applied
twice: every quadratic vertex accepts a mixed-helicity pair, and the three
cyclic choices specify which two incoming legs first form the on-shell
intermediate graviton.  The all-equal configurations fail this test in
every channel.

%---------------------------------------------------------------------
\subsection{Mixed-helicity components of the matrix element}
\label{app:subsec:three_to_one_mixed_channels}
%---------------------------------------------------------------------

We now evaluate the mixed incoming-helicity channels.  Up to permutations
of the three incoming legs, the only two independent families are
\begin{equation}
 (++-)
 \qquad\hbox{and}\qquad
 (--+).
 \label{app:eq:two_mixed_helicity_families}
\end{equation}

First consider the $\delta T_2$ contribution.  Let
$\mathfrak a_{\lambda_4}(4;i|j,k)$ denote the contribution obtained from
the complete ordered all-sum kernel when $i$ occupies the linear slot and $(j,k)$
occupy the pair-symmetrized quadratic slot.  In the
normalization of Eq.~\eqref{eq:compact_integration_notation}, it is
\begin{equation}
\boxed{
\begin{aligned}
\mathfrak a_{\lambda_4}(4;i|j,k)
={}&
-\frac{\mathcal N^{(3)}(4;i,j,k)}{(2\pi)^6}
\left.
\mathscr P_{\lambda_4,z}
\left\{
 \bigl[l(z)\cdot\widehat{k}_4\bigr]
 \mathfrak C_{R,++}^{(3),\mathrm{work}}(z;i|j,k)
\right\}
\right|_{z=\widehat{k}_4},
\\
\mathscr P_{+,z}&=\eth_z^2,
\qquad
\mathscr P_{-,z}=\bar\eth_z^2,
\end{aligned}
}
\label{app:eq:ordered_deltaT2_mixed_kernel}
\end{equation}
where
\begin{equation}
\begin{aligned}
\mathfrak C_{R,++}^{(3),\mathrm{work}}(z;i|j,k)
={}&
\left[
 \mathcal T_3^{(IV)}
 +\mathcal T_{3,\Omega}^{(++)}
 -\mathcal I_C^{(++)}
\right](z;i|j,k)_{\mathrm{pair}}.
\end{aligned}
\label{app:eq:ordered_complete_QSII_source}
\end{equation}
The subscript ``pair'' means the average under $j\leftrightarrow k$.  The
quantity $\mathcal N^{(3)}$ is the complete three-mode normalization defined
with the cubic radiative vertex in Ref.~\cite{QSII}.  The
factor $(2\pi)^{-6}$ converts the three measures and the spatial delta used
in Ref.~\cite{QSII} to the convention of
Eq.~\eqref{eq:compact_integration_notation}.  Equations
\eqref{app:eq:ordered_deltaT2_mixed_kernel} and
\eqref{app:eq:ordered_complete_QSII_source} contain the complete ordered
cone terms and the subtraction of the cubic BCH iteration.  No
recursive-cut contribution is present because
\(U=\mathrm{Id}\).

With the same no-double-counting convention used for the quadratic Bose
average in Ref.~\cite{QSI}, the fully symmetric coefficient is
\begin{equation}
 \alpha_{\lambda_4;\lambda_1\lambda_2\lambda_3}(4;1,2,3)
 =
 \frac13
 \sum_{\mathrm{cyc}(1,2,3)}
 \mathfrak a_{\lambda_4}(4;1|2,3).
 \label{app:eq:alpha_from_ordered_QSII_kernel}
\end{equation}
The explicit ordered-kernel evaluation shows that the contribution whose
quadratic pair has equal helicities vanishes.  Therefore
Eq.~\eqref{app:eq:alpha_from_ordered_QSII_kernel} reduces to
\begin{equation}
\boxed{
\begin{aligned}
\alpha_{\lambda_4;++-}(4;1,2,3)
={}&\frac13\Big[
\mathfrak a_{\lambda_4}
 (4;1_{+}|2_{+},3_{-})
\\
&\hspace{17mm}+
\mathfrak a_{\lambda_4}
 (4;2_{+}|3_{-},1_{+})
\Big],
\\[2mm]
\alpha_{\lambda_4;--+}(4;1,2,3)
={}&\frac13\Big[
\mathfrak a_{\lambda_4}
 (4;1_{-}|2_{-},3_{+})
\\
&\hspace{17mm}+
\mathfrak a_{\lambda_4}
 (4;2_{-}|3_{+},1_{-})
\Big],
\qquad \lambda_4=\pm.
\end{aligned}
}
\label{app:eq:alpha_mixed_two_terms}
\end{equation}
This two-term reduction follows from the complete ordered kernel, rather
than from the raw cone terms alone.  Since no recursive-cut contribution is
present for \(U=\mathbf1\), the mixed-helicity coefficients are determined
by Eq.~\eqref{app:eq:ordered_complete_QSII_source}, including the
cancellations between the ordered cone terms and
\(-\mathcal I_C^{(++)}\).

The contracted contribution can be written without any remaining momentum or
helicity sum.  Define, for any pair of incoming legs,
\begin{equation}
 r_{ij}
 =
 \left(
 |\vec k_i+\vec k_j|,
 \vec k_i+\vec k_j
 \right),
 \qquad
 \omega_{ij}=|\vec k_i+\vec k_j|.
 \label{app:eq:rij_general_definition}
\end{equation}
For $(\lambda_1,\lambda_2,\lambda_3)=(+,+,-)$, the $312$ channel
vanishes, whereas the $123$ and $231$ channels have internal helicity $-$.
Substitution in Eq.~\eqref{app:eq:M_three_to_one_kernel_result} gives
\begin{equation}
\boxed{
\begin{aligned}
&\mathscr M^{(2)}_{\lambda_4;++-}(4;1,2,3)
\\
&\quad=
\delta^{(3)}
(\vec k_4-\vec k_1-\vec k_2-\vec k_3)
\Bigg\{
\frac34
\alpha_{\lambda_4;++-}(4;1,2,3)
\\
&\qquad\quad+
\frac14
\frac{
\kappa^{aa}_{\lambda_4;+-}(4;1,r_{23})
\kappa^{aa}_{-;+-}(r_{23};2,3)
}{2\omega_{23}}
\\
&\qquad\quad+
\frac14
\frac{
\kappa^{aa}_{\lambda_4;+-}(4;2,r_{31})
\kappa^{aa}_{-;-+}(r_{31};3,1)
}{2\omega_{31}}
\Bigg\},
\qquad
\lambda_4=\pm.
\end{aligned}
}
\label{app:eq:M_three_to_one_mixed_explicit}
\end{equation}

For $(\lambda_1,\lambda_2,\lambda_3)=(-,-,+)$ the same two channels
are nonvanishing, now with internal helicity $+$.  The result is
\begin{equation}
\boxed{
\begin{aligned}
&\mathscr M^{(2)}_{\lambda_4;--+}(4;1,2,3)
\\
&\quad=
\delta^{(3)}
(\vec k_4-\vec k_1-\vec k_2-\vec k_3)
\Bigg\{
\frac34
\alpha_{\lambda_4;--+}(4;1,2,3)
\\
&\qquad\quad+
\frac14
\frac{
\kappa^{aa}_{\lambda_4;-+}(4;1,r_{23})
\kappa^{aa}_{+;-+}(r_{23};2,3)
}{2\omega_{23}}
\\
&\qquad\quad+
\frac14
\frac{
\kappa^{aa}_{\lambda_4;-+}(4;2,r_{31})
\kappa^{aa}_{+;+-}(r_{31};3,1)
}{2\omega_{31}}
\Bigg\},
\qquad
\lambda_4=\pm.
\end{aligned}
}
\label{app:eq:M_three_to_one_mixed_conjugate_explicit}
\end{equation}
Every quadratic kernel in Eq.~\eqref{app:eq:M_three_to_one_mixed_explicit}
and Eq.~\eqref{app:eq:M_three_to_one_mixed_conjugate_explicit} is one of
the nonvanishing mixed-helicity $2\to1$ kernels calculated in
Ref.~\cite{QSI}; every $\delta T_2$ term in
Eq.~\eqref{app:eq:alpha_mixed_two_terms} is obtained from the quantum
kernel in Eq.~\eqref{app:eq:ordered_complete_QSII_source}.

The remaining four mixed orderings require no new integrations.  Bose
symmetry gives
\begin{align}
\mathscr M^{(2)}_{\lambda_4;+-+}(4;1,2,3)
&=
\mathscr M^{(2)}_{\lambda_4;++-}(4;1,3,2),
\label{app:eq:M_mixed_permutation_one}
\\
\mathscr M^{(2)}_{\lambda_4;-++}(4;1,2,3)
&=
\mathscr M^{(2)}_{\lambda_4;++-}(4;2,3,1),
\label{app:eq:M_mixed_permutation_two}
\\
\mathscr M^{(2)}_{\lambda_4;-+-}(4;1,2,3)
&=
\mathscr M^{(2)}_{\lambda_4;--+}(4;1,3,2),
\label{app:eq:M_mixed_permutation_three}
\\
\mathscr M^{(2)}_{\lambda_4;+--}(4;1,2,3)
&=
\mathscr M^{(2)}_{\lambda_4;--+}(4;2,3,1).
\label{app:eq:M_mixed_permutation_four}
\end{align}
These relations cover all six mixed incoming-helicity assignments for each
outgoing helicity.  The two all-equal assignments vanish as established in
Eq.~\eqref{eq:main_three_to_one_equal_helicity_zero}.

%---------------------------------------------------------------------

%=====================================================================
\section{Construction of the \texorpdfstring{$2\to2$}{2 to 2} kernel
by signed crossing and helicity selection rules}
\label{app:two_to_two_crossing}
%=====================================================================

This appendix constructs the $2\to2$ kernel by crossing one incoming leg of
the $3\to1$ result and verifies the construction in the explicit
third-order NSF cone kernels.  The calculation determines the point kernel,
the crossed tree-level contraction channels, and the generic helicity
selection rules of the $2\to2$ matrix element.
By ``signed crossing'' we mean the replacement
\(\operatorname{Cr}_3\) defined below, applied consistently to the signed
Fourier momenta, operator monomials, momentum distributions, and cone
kernels.  This construction converts one incoming leg of the \(3\to1\)
coefficient into an outgoing leg while keeping all physical one-particle
energies positive.

\subsection{Signed crossing of one Fourier mode}
\label{appC:subsec:signed_crossing}

All physical one-particle labels retain positive energy
$\omega_i=|\vec k_i|$.  Crossing is implemented in the signed Fourier
coefficient, not by extending the on-shell measure to negative energies.
For leg $3$ define
\begin{equation}
\boxed{
\operatorname{Cr}_3:
\qquad
\eta_3=+1\longmapsto-1,
\qquad
a_{\lambda_3}(k_3)\longmapsto
a^\dagger_{-\lambda_3}(k_3).
}
\label{appC:eq:crossing_definition}
\end{equation}
Here $\eta_3k_3$ is the signed momentum that occurs in the Fourier
derivatives and affine denominators.  The positive-energy measure and the
celestial direction of the physical outgoing graviton are unchanged.

The all-annihilator branch contains
\begin{equation}
(2\pi)^3\delta^{(3)}
\!\left(\vec k_4-\vec k_1-\vec k_2-\vec k_3\right)
a_{\lambda_1}(1)a_{\lambda_2}(2)a_{\lambda_3}(3).
\label{appC:eq:all_annihilator_branch}
\end{equation}
Crossing the third leg gives
\begin{align}
&(2\pi)^3\delta^{(3)}
\!\left(\vec k_4-\vec k_1-\vec k_2+\vec k_3\right)
a^\dagger_{-\lambda_3}(3)a_{\lambda_1}(1)a_{\lambda_2}(2)
\nonumber\\
&\qquad=
(2\pi)^3\delta^{(3)}
\!\left(\vec k_1+\vec k_2-\vec k_3-\vec k_4\right)
a^\dagger_{-\lambda_3}(3)a_{\lambda_1}(1)a_{\lambda_2}(2).
\label{appC:eq:crossed_operator_and_delta}
\end{align}
This is precisely the mixed normal-ordered branch
$a_3^\dagger a_1a_2$.  Thus putting one incoming signed momentum in the
outgoing direction converts the $3\to1$ spatial support into the $2\to2$
support.

\subsection{Crossing of the cone form factors}
\label{appC:subsec:kernel_crossing}

The signed momentum of the quadratic sub-block is
\begin{equation}
K^{(2)}_{\eta_2\eta_3}=\eta_2k_2+\eta_3k_3,
\qquad
K^{(2)}_{++}\xrightarrow{\operatorname{Cr}_3}K^{(2)}_{+-}=k_2-k_3.
\label{appC:eq:K_sum_to_difference}
\end{equation}
The conformal-factor form factor transforms explicitly as
\begin{align}
\operatorname{Cr}_3F_\Omega^{(+,+)}
={}&
-\bigl[l^+(z)\cdot k_2\bigr]
 \bigl[l^+(z)\cdot k_3\bigr]
\nonumber\\
&+
\frac{
 \bigl[l^+(z)\cdot k_2\bigr]^2
 \bigl[l^+(z)\cdot k_3\bigr]^2
}{
 \bigl[l^+(z)\cdot(k_2-k_3)\bigr]^2
}
=F_\Omega^{(+,-)}.
\label{appC:eq:F_Omega_crossing}
\end{align}
Similarly,
\begin{equation}
\operatorname{Cr}_3F_\Lambda^{(+,+)}
=-(k_2\cdot k_3)
=F_\Lambda^{(+,-)}.
\label{appC:eq:F_Lambda_crossing}
\end{equation}
The factors of $\ii$ generated by the Fourier derivatives are essential for
these relative signs.

The affine momentum and the two numerators in the mixed angular-field term
obey
\begin{align}
k_1+k_2+k_3&\longmapsto k_1+k_2-k_3,
\nonumber\\
k_1\cdot(k_2+k_3)&\longmapsto k_1\cdot(k_2-k_3),
\nonumber\\
(k_1+k_2)\cdot k_3&\longmapsto-(k_1+k_2)\cdot k_3.
\label{appC:eq:mixed_source_crossing}
\end{align}
Together with
$\mathcal H_{\rm sum}\mapsto\mathcal H_{\rm diff}$, substitution in the
explicit kernels of Ref.~\cite{QSII} gives
\begin{equation}
\boxed{
\operatorname{Cr}_3\mathcal T_3^{(IV)}=\mathcal T_3^{(III)},
\qquad
\operatorname{Cr}_3\mathcal T_{3,\Omega}^{(++)}
=\mathcal T_{3,\Omega}^{(+-)}.
}
\label{appC:eq:T3_branch_crossing}
\end{equation}
The linear-metric cone term and the BCH kernel transform by the same
signed-mode rule.  Hence, within the cone
truncation used to reconstruct $\delta T_2$,
\begin{equation}
\boxed{
\operatorname{Cr}_3\mathcal C_{R,++}^{(3),\rm cone}
=\mathcal C_{R,+-}^{(3),\rm cone}.
}
\label{appC:eq:C_cone_branch_crossing}
\end{equation}
No recursive-cut contribution is present because \(U=\mathbf1\).

Including the common mode normalization and the radiative projector, the
complete distributional relation is
\begin{equation}
\boxed{
\begin{aligned}
&(2\pi)^3\delta^{(3)}
\!\left(\vec k_4+\vec k_3-\vec k_1-\vec k_2\right)
\mathcal B_{R,\lambda_4\lambda_3;\lambda_1\lambda_2}(4,3|1,2)
\\
&\quad=
\operatorname{Cr}_3\!\left[
(2\pi)^3\delta^{(3)}
\!\left(\vec k_4-\vec k_1-\vec k_2-\vec k_3\right)
\mathcal A_{R,\lambda_4;\lambda_1\lambda_2,-\lambda_3}(4;1,2,3)
\right].
\end{aligned}
}
\label{appC:eq:full_B_as_crossed_A}
\end{equation}
After stripping the identical spatial distributions, while retaining the
mode normalization in the coefficient kernels,
\begin{equation}
\boxed{
\mathcal B_{R,\lambda_4\lambda_3;\lambda_1\lambda_2}(4,3|1,2)
=
\operatorname{Cr}_3
\mathcal A_{R,\lambda_4;\lambda_1\lambda_2,-\lambda_3}(4;1,2,3).
}
\label{appC:eq:B_as_crossed_A}
\end{equation}

\subsection{Point vertex and crossed canonical contractions}
\label{appC:subsec:point_and_contractions}

The two creators in Eq.~\eqref{eq:two_to_two_iT2_sector} contract with the
two outgoing annihilators in $2!$ ways, and the two annihilators contract
with the incoming creators in $2!$ ways.  Therefore
\begin{equation}
(2!)(2!)\left(\frac12\right)^4=\frac14,
\label{appC:eq:point_external_multiplicity}
\end{equation}
which proves Eq.~\eqref{eq:two_to_two_point_matrix_element}.  Equivalently,
the already calculated decay kernel gives
\begin{equation}
\boxed{
\begin{aligned}
\mathscr M^{(2),\rm point}_{
\lambda_3\lambda_4;\lambda_1\lambda_2}
={}&\frac14(2\pi)^3\delta^{(3)}
\!\left(\vec k_1+\vec k_2-\vec k_3-\vec k_4\right)
\\
&\times
\left[
\operatorname{Cr}_3
\mathcal A_{R,\lambda_4;\lambda_1\lambda_2,-\lambda_3}(4;1,2,3)
\right]_{(34)(12)}.
\end{aligned}
}
\label{appC:eq:point_amplitude_from_decay_kernel}
\end{equation}

Write the two-annihilator part of the cubic generator as
\begin{equation}
\mathcal T_1=-2\ii\int_{pij}
\mathcal K^{aa}(p;i,j)a^\dagger(p)a(i)a(j),
\qquad
\delta T_1=\mathcal T_1+\mathcal T_1^\dagger.
\label{appC:eq:T1_for_two_to_two}
\end{equation}
The decay calculation contains two factors $\mathcal T_1$.  Crossing one
annihilation leg converts one of them into $\mathcal T_1^\dagger$.  The
crossing preserves the contraction multiplicities but changes the product
of vertex coefficients by
\begin{equation}
\frac{(-2\ii)(+2\ii)}{(-2\ii)(-2\ii)}=-1,
\qquad
+\frac14\longmapsto-\frac14.
\label{appC:eq:contraction_coefficient}
\end{equation}
The three cyclic decay contraction channels therefore become
\begin{equation}
\boxed{
\begin{aligned}
\mathcal E_s&=\operatorname{Cr}_3\mathcal D(4;3|1,2),
&\vec q_s&=\vec k_1+\vec k_2,
\\
\mathcal E_t&=\operatorname{Cr}_3\mathcal D(4;2|3,1),
&\vec q_t&=\vec k_1-\vec k_3,
\\
\mathcal E_u&=\operatorname{Cr}_3\mathcal D(4;1|2,3),
&\vec q_u&=\vec k_1-\vec k_4.
\end{aligned}
}
\label{appC:eq:three_crossed_tree_channels}
\end{equation}
The spatial deltas fix the internal momentum in each contraction channel, so no loop
integration remains in these tree terms.

\subsection{Helicity selection rules}
\label{appC:subsec:selection_rules}

For generic non-collinear data, the explicit three-annihilator-kernel
calculation gives
\begin{equation}
\mathcal A_{R,\lambda_4;+++}=0,
\qquad
\mathcal A_{R,\lambda_4;---}=0,
\qquad \lambda_4=\pm.
\label{appC:eq:A_all_equal_zero_recalled}
\end{equation}
Let $\lambda_1=\lambda_2=\sigma$.  Since the decay helicity of the crossed
leg is $-\lambda_3$, Eq.~\eqref{appC:eq:B_as_crossed_A} gives
\begin{equation}
\mathcal B_{R,\lambda_4\lambda_3;\sigma\sigma}=0
\qquad\text{if}\qquad \lambda_3=-\sigma.
\label{appC:eq:B_zero_first_outgoing_leg}
\end{equation}
Creator symmetry gives the companion condition
\begin{equation}
\mathcal B_{R,\lambda_4\lambda_3;\sigma\sigma}=0
\qquad\text{if}\qquad \lambda_4=-\sigma.
\label{appC:eq:B_zero_second_outgoing_leg}
\end{equation}
Thus an equal-helicity incoming pair can produce only the same
equal-helicity outgoing pair.  For a mixed incoming pair and an
equal-helicity outgoing pair, the crossed Hermiticity relation
\begin{equation}
\mathcal B_R(3,4|1,2)
=-\left[\mathcal B_R(1,2|3,4)\right]^*
\label{appC:eq:B_reverse_relation}
\end{equation}
reduces the statement to the reverse process, which vanishes by
Eqs.~\eqref{appC:eq:B_zero_first_outgoing_leg} and
\eqref{appC:eq:B_zero_second_outgoing_leg}.  The resulting generic selection
rules are
\begin{equation}
\begin{array}{c|c}
\text{allowed by the selection rules}&\text{zero on generic support}
\\ \hline
++\to++,\quad --\to--
&++\to+-,\quad ++\to-+,\quad ++\to--
\\
+-\to+-,\quad +-\to-+
&+-\to++,\quad +-\to--
\end{array}
\label{appC:eq:two_to_two_selection_table}
\end{equation}

Channels related by permutations of identical bosonic external legs are not
listed separately in Eq.~\eqref{appC:eq:two_to_two_selection_table}.

\end{document}